\documentclass[aps, twocolumn, prl, amssymb, amsmath, superscriptaddress]{revtex4-2}
\usepackage{graphicx}
\usepackage{dcolumn}
\usepackage{bm}
\usepackage{amsmath}
\usepackage{subfigure}
\usepackage{amsfonts}
\usepackage[svgnames]{xcolor}
\usepackage{appendix}
\usepackage{multirow}
\usepackage{booktabs}
\usepackage{tabularx}
\usepackage[colorlinks,
            linkcolor=blue,
			filecolor=blue,
			urlcolor= NavyBlue,
			citecolor=blue,
            ]{hyperref}
\usepackage{amsthm}
\usepackage{setspace}
\usepackage[T1]{fontenc}
\usepackage{lineno}

\usepackage[percent]{overpic}
\usepackage{dsfont}
\usepackage{floatrow}
\usepackage{physics}

\newcommand{\kett}[1]{\left.\ket{#1}\right\rangle}

\newcommand{\brakett}[2]{\left\langle\braket{#1}{#2}\right\rangle}
\newcommand{\ee}{{e}}
\newcommand{\ii}{{i}}
\newcommand{\phii}[0]{\hat{\Phi}}
\newcommand{\mm}[0]{\hat{\mathcal{M}}}
\newcommand{\uu}[0]{\hat{\mathcal{U}}}

\begin{document}

\preprint{APS/123-QED}

\title{Theory of Metastability in Discrete-Time Open Quantum Dynamics}
\author{Yuan-De Jin}\thanks{These authors contributed equally to this work}
\affiliation{Department of Applied Physics, University of Science and Technology Beijing, Beijing 100083, China}
\author{Chu-Dan Qiu}\thanks{These authors contributed equally to this work}
\affiliation{State Key Laboratory of Superlattices and Microstructures, Institute of Semiconductors, Chinese Academy of Sciences, Beijing, 100083, China}
\author{Wen-Long Ma}
\email{wenlongma@semi.ac.cn}
\affiliation{State Key Laboratory of Superlattices and Microstructures, Institute of Semiconductors, Chinese Academy of Sciences, Beijing, 100083, China}
\affiliation{Center of Materials Science and Opto-Electronic Technology,\\ University of Chinese Academy of Sciences, Beijing 100049, China}

\date{\today}

\begin{abstract}
 Metastability in open system dynamics describes the phenomena of initial relaxation to long-lived metastable states before decaying to the asymptotic stable states. It has been predicted in continuous-time stochastic dynamics of both classical and quantum systems. Here we present a general theory of metastability in discrete-time open quantum dynamics, described by sequential quantum channels. We focus on a general class of quantum channels on a target system, induced by an ancilla system with a pure-dephasing coupling to the target system and under Ramsey sequences. Interesting metastable behaviors are predicted and numerically demonstrated by decomposing the average dynamics into stochastic trajectories. Examples and applications are also discussed.
\end{abstract}

\maketitle


\textit{Introduction.---}Metastability, ubiquitous in open systems, arises when the system dynamics settle into long-lived states before ultimately decaying to true equilibria.  In classical stochastic dynamics, metastability emerges when there is a separation of time scales due to a spectrum splitting in the generator of the master equations \cite{B.Gaveau_1987,cheng1998,Bernard.Gaveau_2000,gaveauMultiplePhasesStochastic2006b}. Previous studies have investigated metastability in many systems, such as phase transitions in finite-size systems \cite{gaveauTheoryNonequilibriumFirstorder1998a,GAVEAU1999222PT,huisinga2004,boettger1997,anderson2018} and non-equilibrium dynamics of quantum many-body systems \cite{binder1986,jckle1991,sollich1999,cugliandolo1999,garrahan2002,sollich2003,biroli2013,lan2018}.

By extending metastability from classical stochastic dynamics to the quantum domain, quantum metastability theory has been formulated for continuous-time Markovian dynamics \cite{PhysRevLett.116.240404,Merkli_2015,macieszczak2021,macieszczak2021a,brown2023a}, described mostly by Lindblad master equations \cite{lindblad1976generators,gorini1976}. Manifold of metastable states is argued to be composed of disjoint states, decoherence-free subspaces and noiseless subsystems \cite{PhysRevLett.116.240404}. With this theoretical framework, quantum metastability has been found in various settings, such as dissipative phase transitions of the quantum Ising model \cite{rose2016}, dynamics of quantum systems coupling with dissipative bosonic modes \cite{PhysRevLett.129.063601}, the driven-dissipative setting of Rabi model \cite{leboite2017}, Bose-Hubbard model \cite{liu2022} and long-range interacting systems \cite{doi:10.1073/pnas.2101785118}, and experiments of Rydberg gases \cite{letscher2017}. Other novel phenomena include metastable discrete time-crystalline phases in Floquet open systems \cite{Gong2018,Gambetta2019,cabot2022} and Majorana bosons in metastable quadratic Markovian dynamics \cite{flynn2021}.

\begin{figure}
\centering
  \includegraphics[width=8cm]{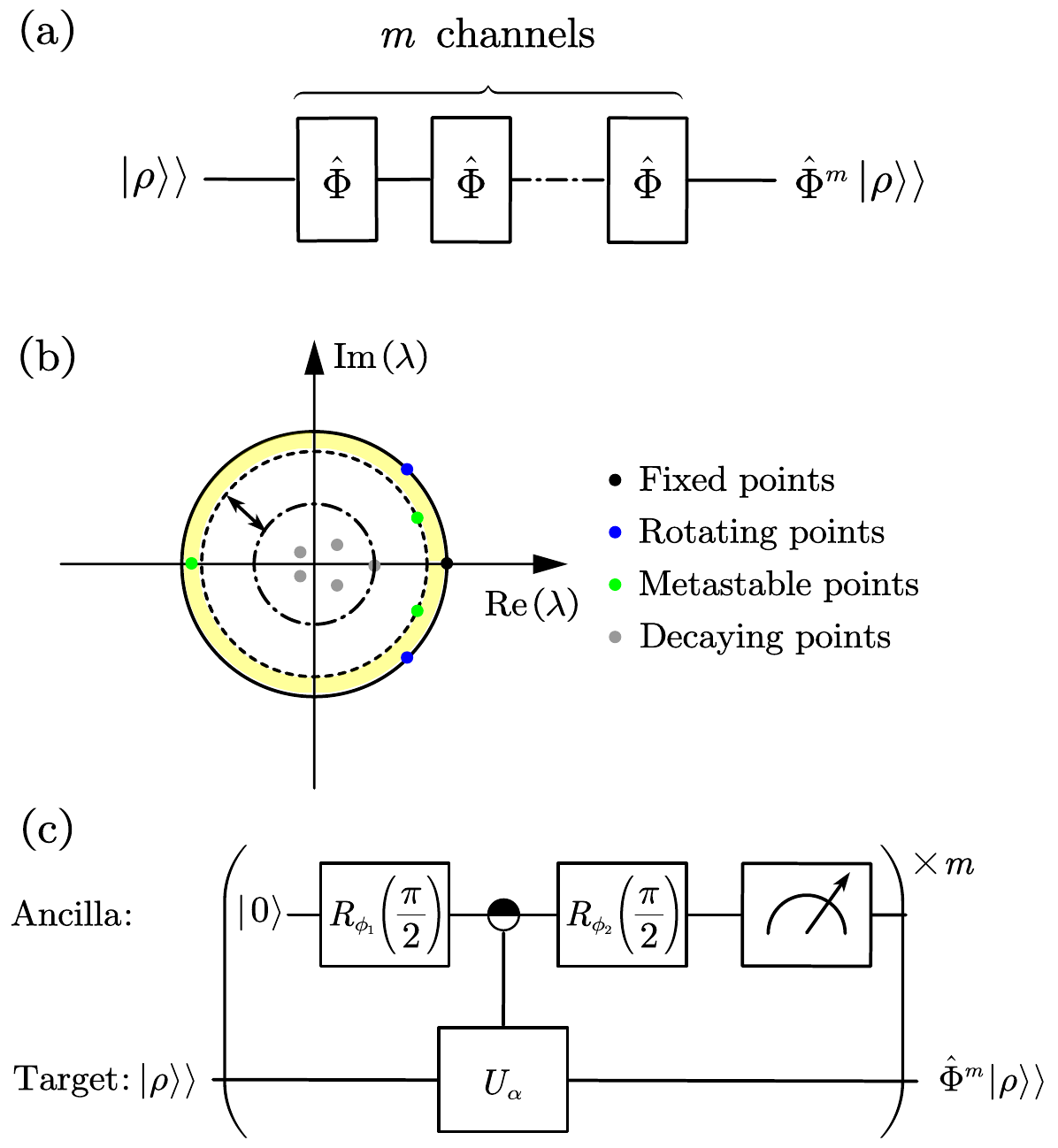}
  \caption{(a) Schematic of sequential quantum channels. For an initial state $\kett\rho$ of a target system, its final state becomes $\hat{\Phi}^m|\rho\rangle\rangle$ after applying the quantum channel for $m$ times. (b) The eigenvectors of a quantum channel can be divided into three categories: the fixed points with eigenvalue $\lambda=1$, rotating points with $\lambda=e^{i\varphi}$ $(\varphi\neq 0)$, and decaying points with $|\lambda|<1$. The decaying points with $|\lambda|\approx 1$ are also called metastable points. Note that the complex eigenvalues of a quantum channel always come in conjugate pairs. The area shaded in yellow labels a metastable region and quantum metastability emerges when the gap (represented by an arrow) between the smallest eigenvalue in this region and the next smaller one outside the region is relatively large.
  (c) Schematic of the quantum circuit for sequential RIMs, where $R_{\phi}(\theta)=\ee^{-i (\cos\phi\sigma_q^x+\sin \phi\sigma_q^y)\theta /2}$ is the ancilla rotation operator and $U_{\alpha}=e^{-i[(-1)^{\alpha}B+\gamma C]}$ is a unitary operator of the target system conditioned on the ancilla state $\ket{\alpha}_q$ ($\alpha=0,1$).}
  \label{fig1}
\end{figure}

However, current works on quantum metastability mainly concern continuous-time open quantum dynamics. It remains largely unexplored whether similar phenomena can occur for discrete-time open quantum dynamics, which can be described by sequential quantum channels [see Fig. \ref{fig1}{\color{blue}(a)}] (also called discrete-time quantum Markov chains) \cite{gudder2008,guan2018,novotny2018quantum,amato2023}. Sequential quantum channels appear in a broad range of scenarios, such as quantum random walks \cite{attal2012}, quantum collision models \cite{ciccarello2022} and quantum channel simulations \cite{lloyd2001, andersson2008, shen2017, han2021,cai2021}.

In this paper, we formulate a general theory of metastability in sequential quantum channels and derive the conditions for observing metastability. The theory is based on spectrally decomposing a quantum channel and classifying its eigenvalues. The key finding is to predict interesting metastability behaviors for sequential quantum channels on a target system induced by sequential Ramsey interferometry measurements (RIMs) of an ancilla system, which is a common protocol in quantum information processing \cite{degen2017}. We confirm the theoretical analysis by decomposing the average dynamics of sequential channels into stochastic trajectories with Monte Carlo simulations of practical examples. The findings also provide a theoretical foundation for recent experiments in polarizing a quantum environment with an ancilla qubit in solid-state systems \cite{dasari2022,madzik2020,liu2017,bhaktavatsalarao2019}.

\textit{Model for metastability in sequential quantum channels.---}We first present a general model for quantum metastability in sequential quantum channels, which can be regarded as a generalization of the continuous-time framework in \cite{PhysRevLett.116.240404} to the discrete-time case. However,
the extension is non-trivial since the channel in each cycle is not necessarily the integration of a Lindblad generator within a small time step and can be qubite arbitrary (e.g., highly non-Markovian).

Quantum channel is a completely positive and trace-preserving (CPTP) map \cite{kraus1983states,caruso2014quantum,wolf2011url,watrous2018theory}, which maps a density operator to another by $\Phi(\rho)=\rho'$. One can represent a quantum channel in the Kraus representation as \cite{kraus1983states}
\begin{equation}
  \Phi(\cdot)=\sum_\alpha M_{\alpha} (\cdot) M_{\alpha}^\dagger=\sum_\alpha \mathcal{M}_{\alpha}(\cdot),
\end{equation}
where $\{M_{\alpha}\}$ are a set of Kraus operators satisfying $\sum_\alpha M^{\dagger}_{\alpha} M_{\alpha}=\mathbb{I}$ with $(\cdot)^\dagger$ denoting the Hermitian conjugation and $\mathbb{I}$ being the identity operator, and $\mathcal{M_{\alpha}}(\cdot)=M_{\alpha} (\cdot) M_{\alpha}^{\dagger}$ is a superoperator. The set of operators $\{M_{\alpha}^{\dagger}M_{\alpha}\}$ form a positive operator-valued measure (POVM) representing a generalized measurement, which can also be simulated by projective measurements and postselection \cite{oszmaniec2017,oszmaniec2019,singal2022,linden2023}.

Quantum channel also has a natural representation in the Hilbert-Schmidt (HS) space \cite{watrous2018theory,bengtsson2017geometry,Note1}. A linear operator on a Hilbert space is transformed to a ket in the HS space $A=\sum_{ij}a_{ij}\ket{i} \bra{j}$$\leftrightarrow$$\kett{A}=\sum_{ij}a_{ij}\kett{ij}$, and the inner product in HS space is defined as $\brakett{A}{B}=\Tr(A^\dagger B)$. Then a superoperator on Hilbert space corresponds to a linear operator on HS space: $X(\cdot)Y\rightarrow X\otimes Y^T\kett{\cdot}$, so that $\hat{\mathcal{M}}_\alpha=M_\alpha\otimes M^*_\alpha$ and $\hat\Phi=\sum_\alpha \hat{\mathcal{M}}_\alpha$, where $(\cdot)^T$ and $(\cdot)^*$ denote the matrix transposition and matrix conjugation, respectively. Note that we add hats on operators acting on HS space.

Since the natural representation of a quantum channel is a linear operator on the HS space, it can be spectrally decomposed as \cite{wolf2011url}
\begin{equation}
  \hat\Phi=\sum_i \lambda_i |R_i\rangle\rangle \langle\langle L_i|,
\end{equation}
where $\lambda_i=\abs{\lambda_i}\ee^{i\varphi_i}$ is the $i$th eigenvalue and $\kett{R_i}$($\kett{L_i}$) is the corresponding right (left) eigenvector, satisfying $\hat \Phi\kett{R_i}=\lambda_i \kett{R_i}$, $\hat \Phi^\dagger\kett{L_i}=\lambda_i^{*} \kett{L_i}$, and the biorthonormalization condition $\brakett{L_i}{R_j}=\mathrm{Tr}(L_i^\dagger R_j)=\delta_{ij}$. Here we assume that the channel is diagonalizable (see \footnote{See Supplemental Material includes Ref. \cite{wolf2011url,watrous2018theory,PhysRevLett.116.240404,ma2016} at \url{} for more details.} for a general Jordan decomposition of a channel). The eigenvalues $\{\lambda_i\}$ of a quantum channel are all located within a unit disk of the complex plane \cite{kraus1983states}, and we order them by decreasing modulus, $\abs{\lambda_{i}}\geq \abs{\lambda_{i+1}}$ [Fig. \ref{fig1}{\color{blue}(b)}]. The eigenvectors with eigenvalue 1 are called \textit{fixed points} \cite{arias2002fixed,Burgarth2013} denoted as $\kett{\rho_{\rm fix}^i}$, those with eigenvalue $\ee^{\ii\varphi}$ ($\varphi\neq 0$) are \textit{rotating points} \cite{Albert2019asymptoticsof}, {\color{black}and those with $\abs{\lambda_i}<1$ are \textit{decaying points}}. The state subspace spanned by the fixed points and rotating points are asymptotic subspace (also known as peripheral or attractor subspace). The decaying points with eigenvalue $|\lambda_i|\approx 1$ are called \textit{metastable points}.

For a channel $\hat\Phi$ with $n$ fixed points, quantum metastability can emerge when there are $l-n$ metastable points. After sequentially applying the quantum channel for $m$ times, we have \footnote{Here we assume that the channel has no rotating points, see \cite{Note1} for the form of sequential channels that have rotating points.}
\begin{equation}
\begin{aligned}\label{3}
	    \hat\Phi^m\kett{\rho}\simeq\sum_{i=1}^n c_i\kett{\rho_{\rm fix}^i}+\sum_{j=n+1}^{l}c_j\ee^{ m(\ln{\abs{\lambda_j}+\ii \varphi_j)}} \kett{R_j},
\end{aligned}
\end{equation}
with $\lambda_j=\abs{\lambda_j}\ee^{\ii\varphi_j}$ and $c_j={\Tr}(L_j^\dagger \rho)$. The contribution of the other decaying points decays fast as $m$ grows, and can be omitted when $m\gg \mu''=1/\abs{\ln\abs{\lambda_{l+1}}}$ \cite{PhysRevLett.116.240404}. The metastable points cannot be neglected when $m\ll \mu' = 1/\abs{\ln \abs{\lambda_l}}$. So $\mu'$ and $\mu''$ delimit a metastable region: $\mu''\ll m\ll \mu'$, where the metastable points with real eigenvalues act like fixed points, and those with complex eigenvalues act like rotating points.

The eigenvalues of a quantum channel can appear in conjugate pairs, i.e., for an eigenvalue $\lambda_{j,1}=\abs{\lambda_j}\ee^{\ii\varphi_j}$, we have $\lambda_{j,2}=\abs{\lambda_j}\ee^{-i\varphi_j}$. Then we let $c_{j,1}'=\abs{c_{j,1}}\cos(m\varphi_j+\delta_j),c_{j,2}'=\abs{c_{j,2}}\sin(m\varphi_j+\delta_j),|{R_{j,1}'}\rangle\rangle=\kett{R_{j,1}}+\kett{R_{j,2}}$ and $|{R_{j,2}'}\rangle\rangle=i(\kett{R_{j,1}}-\kett{R_{j,2}})$ with $\delta_j={\rm arg}(c_{j,1})$, so Eq. (\ref{3}) becomes
\begin{equation}\label{channelmeta}
   \hat\Phi^m\kett{\rho}\simeq\sum_{i=1}^n c_i \kett{\rho_{\rm fix}^i}+\sum_{j=n+1}^{l}c_j'(m)\kett{R_j'}.
\end{equation}
For real eigenvalues, $c_j'=c_j$ and $|R_j'\rangle\rangle = \kett{R_j}$, then $\hat\Phi^m\kett{\rho}\simeq\sum_{i=1}^n c_i \kett{\rho_{\rm fix}^i}+\sum_{j=n+1}^{l}c_j\kett{R_j}$, which is independent of $m$. 

A metastable state can be fully determined by $(c_{2},\dots,c_n,c_{n+1}'\dots,c_l')$ \footnote{Considering that $\Tr(R_j)=0$ for decaying points and $\Tr(\rho_{\rm fix}^i)=1$, so $\sum_{i=1}^n c_i+\sum_{j=n+1}^l c'_j=1$, and $c_1=1-\sum_{i=2}^n c_i-\sum_{j=n+1}^l c'_j$}, corresponding to a point in a $(l-1)$-dimensional HS subspace. But $\kett{R'_j}$ is not a physical state since the trace of decaying points is zero \cite{Note1}, i.e., $\Tr(R'_j)=0$. So we should transform the above HS subspace to a metastable manifold (MM), where a metastable state is a convex combination of $l$ disjoint extreme metastable states (EMSs) \cite{gaveauMultiplePhasesStochastic2006b}. Thus $\hat\Phi^m \kett{\rho}\simeq\sum_{v=1}^{l}{p_v}\kett{\rho_v}$
, where $\{\rho_v\}$ is a set of EMSs, and $ p_v=\mathrm{Tr}( P_v\rho)$ satisfying $\sum_{v}  p_v=1$. Here $\{ P_v\}$ is a set of observables satisfying $\langle\langle{ P_v}|{\rho_u}\rangle\rangle=\delta_{vu}$, $ P_v\geq 0$ and $\sum_v { P_v}=\mathbb{I}$. Then sequential channels in the metastable region can be approximated as
\begin{equation}\label{MMChannel}
  \hat\Phi^m \simeq \sum_{v=1}^l \kett{\rho_v}\langle\langle  P_v |.
\end{equation}
 
When $m\gtrsim\mu'$, Eq. (\ref{channelmeta}) no longer holds and the weight of the second term in Eq. (\ref{3}) decreases exponentially as $m$ increases. The system gradually leaks from metastable states and relaxes toward the stationary states corresponding to fixed points. 

\begin{figure*}
\!\!\!\!\includegraphics[width=18cm]{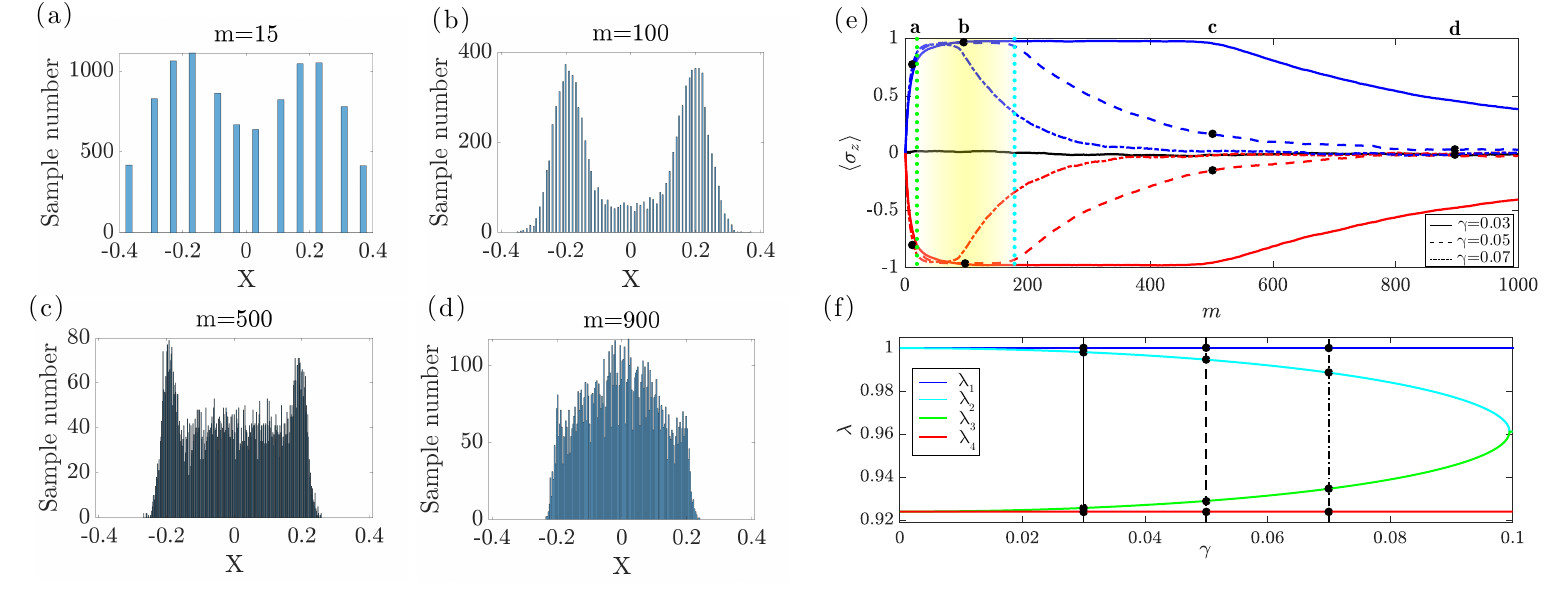}
  \caption{Metastability in measurement statistics of sequential RIMs. (a-d) Monte Carlo simulations for a single target qubit (with $\gamma=0.05$), where we present four stages of evolutions of measurement polarization statistics: (a) When $m$ is relatively small, the width of peaks is large and the two peaks have overlaps; (b) For a larger $m$, quantum metastability emerges and there appears two well-distinguished distribution peaks corresponding to two EMSs; (c) The two peaks gradually disappear as $m$ increases beyond the metastable region; (d) Finally, there appears a single peak corresponding to the maximally mixed state $\mathbb{I}/2$ when $m\gg\mu'$. (e) $\langle{\sigma_z}\rangle$ for three classes of trajectories in Monte Carlo simulations with different $\gamma$: the target qubit is either polarized to nearly $\ket{0}$ (blue lines) or $\ket{1}$ (red lines), or depolarized to $\mathbb{I}/2$ (black lines). The four stages in (a-d) are also labeled with dots. The metastable region is marked with a gradient yellow shade. The boundaries of the region ($\mu''$ and $\mu'$) are labeled with green and cyan dotted line respectively on $m$ axis. (f) Spectrum of the channel as function of $\gamma$, where the spectra corresponding to $\gamma$ in (e) are labeled. All the simulations contains $10^4$ samples with $\Delta\phi=\pi/2$.}
  \label{fig2}
\end{figure*}

\textit{Metastability for sequential RIMs.---}The key result of this paper is to discover quantum metastability in a general class of sequential quantum channels, that is, the channel on a target quantum system induced by an ancilla qubit under RIM sequences [Fig. \ref{fig1}{\color{blue}(c)}]. Suppose that the target system is coupled to an ancilla qubit by a pure-dephasing Hamiltonian as
\begin{equation}\label{hamiltonian}
  H=\sigma_q^z\otimes B+ \gamma \,\mathbb{I}_q\otimes C,
\end{equation}
where $\sigma_q^i$ is the Pauli-$i$ operator of the ancilla, $B$ and $C$ are both operators on the target system and $\gamma$ controls the magnitude of the second term \footnote{In order to use a single parameter $\gamma$ to adjust the relative strength, we assume that $B$ and $C$ have the same operator norm.}.

In a single RIM, an ancilla qubit is initialized to	$\ket{0}_q$ and rotated to $\ket{\psi}_q=R_{\phi_1}(\frac{\pi}{2})\ket{0}_q=(\ket{0}_q-ie^{i\phi_1}\ket{1}_q)/\sqrt{2}$, with the rotation operator being $R_{\phi}(\theta)=e^{-i (\cos\phi\sigma_q^x+\sin \phi\sigma_q^y)\theta /2}$, then interacts with the target system under the Hamiltonian $H$ for time $t$, undergoes another rotation $R_{\phi_2}(\frac{\pi}{2})$ and is finally projectively measured in the basis $\{\ket{0}_q,\ket{1}_q\}$. The measurement result is either $0$ or $1$ for a single RIM.

Such a process induces a quantum channel on the target system, which can be written in Stinespring representation as \cite{stinespring}
\begin{equation}
  \Phi(\rho)=\Tr_q[U(\rho_q\otimes\rho)U^\dagger],
\end{equation}
where \,$\rho_q=\ket{\psi}_q\bra{\psi}$ and $U=e^{-i Ht}=\sum_{\alpha=0,1}|\alpha\rangle_q\langle\alpha|\otimes U_{\alpha}$ with $U_{\alpha}=e^{-i[(-1)^{\alpha} B+\gamma C]}$. By tracing over the ancilla qubit, we get the natural representation
\begin{equation}
  \phii=\mm_0+\mm_1=(\uu_0+\uu_1)/2,\label{RIMKraus}
\end{equation}
where $\mathcal{\hat U_\alpha}=U_\alpha\otimes U_\alpha^*$, and $\mathcal{\hat M_\alpha}=M_\alpha\otimes M_\alpha^*$ with the Kraus operator $M_{\alpha}=[U_0-(-1)^{\alpha} e^{i\Delta\phi}U_1]/2$ and $\Delta\phi=\phi_1-\phi_2$. Note that the channel above depends on the initial state of the ancilla and describes non-Markovian open quantum dynamics.

That the channel induced by RIM [Eq. (\ref{RIMKraus})] is a unital mixed-unitary channel \cite{novotny2010asymptotic,audenaert2008random,watrous2018theory}, satisfying $\Phi(\mathbb{I})=\mathbb{I}$. It has been proven that $\rho$ is a fixed point of a unital channel if and only if it commutes with every Kraus operator \cite{watrous2018theory}, i.e., $[\rho, M_{\alpha}]=0$ for any $\alpha$. When $\gamma\neq0$, the fixed points of the channel [Eq. (\ref{RIMKraus})] depends on the commutativity of $B$ and $C$ (see \cite{Note1} for the proof):

\noindent (i) If $[B,C]=0$, then $B$ and $C$ can be diagonalized simultaneously, $B=\sum_{j=1}^d b_j \ket{j}\bra{j}$ and $C=\sum_{j=1}^d c_j \ket{j}\bra{j}$, so the fixed points are spanned by a set of rank-one projections $\{|j\rangle\langle j|\}_{j=1}^d$, with $d$ denoting the dimension of Hilbert space $\mathcal{H}$ of the target system. 

\noindent(ii) If $[B,C]\neq 0$, $B$ and $C$ can be reduced simultaneously to a block-diagonal form by a unitary transformation $W$, $B=W\left(\bigoplus_{j=1}^r B_j\right) W^{\dagger}$ and $C= W\left(\bigoplus_{j=1}^r C_j\right) W^{\dagger}$, where $W$ is chosen such that $B_j$ and $C_j$ for any $j$ cannot be reduced further to have more blocks. Such a block diagonalization partitions the Hilbert space of the target system $\mathcal{H}$ into the direct sum of $r$ subspaces $\mathcal{H}=\bigoplus_{j=1}^r\mathcal{H}_j$, and $[B_j, {C}_{j}] \neq 0$ for at least one subspace $\mathcal{H}_j$ with ${\rm dim}(\mathcal{H}_j)\geq2$. Then the fixed points are spanned by a set of projection operators $\{\Pi_j\}_{j=1}^r$ ($r\leq d$), where $\Pi_j$ is the projection to the subspace $\mathcal{H}_j$, satisfying $\sum_{j=1}^r\Pi_j=\mathbb{I}$.

Sequential RIMs of the ancilla induces sequential quantum channels on the target system. Thus, for $[B,C]=0$, the asymptotic operation of sequential such channels is a polarizing channel (or a projective measurement) on the target system \cite{bauer2011,haapasalo2016, ma2018phase,PhysRevA.107.012217}, while for $[B, C]\neq0$ it can be a depolarizing channel at least for the subspace $\mathcal{H}_j$ with $\dim(\mathcal{H}_j)\geq2$. Interestingly, if we consider the ancilla dynamics, the target system causes decoherence of the ancilla, and these two cases corresponds exactly to those where the target system produces static thermal or dynamical quantum noise, respectively \cite{yang2017,szankowski2017,liu2012,Reinhard2012}.

As the open system dynamics described by Lindblad master equations can be unraveled with quantum trajectories \cite{plenio1998quantum,brun2002simple}, the average dynamics of sequential quantum channels can also be decomposed into stochastic trajectories \cite{PhysRevA.107.012217}. After repeating RIM for $m$ times, the measurement results of the ancilla will be a sequence of $m$ binary numbers $(\alpha_1,\alpha_2,\cdots,\alpha_m)$ with $\alpha_i\in\{0,1\}$, also inducing an operation $\hat{\mathcal{M}}_{\alpha_m}\cdots,\hat{\mathcal{M}}_{\alpha_1}\hat{\mathcal{M}}_{\alpha_1}$ on the target system (note that $[\hat{\mathcal{M}}_{0}, \hat{\mathcal{M}}_{1}]\neq0$ if $[B,C]\neq0$). Denote the number of $0/1$ is $m_0/m_1$ ($m_0+m_1=m$), one can get a measurement frequency $F=\{m_0/m, m_1/m\}$. The measurement polarization $X=(m_0-m_1)/(2m)$ denotes the different classes of stochastic trajectories that the target system undergoes \cite{ma2018phase,PhysRevA.107.012217}. The measurement distribution of $X$ can show multiple distribution peaks, with each peak corresponding to a fixed point of the channel. Such measurement statistics can be efficiently obtained by Monte Carlo simulations \cite{ma2018phase,Wudarski2023}.

Metastability can occur when $[B,C]\neq 0$ and $\gamma$ is small. When $\gamma=0$, we have shown above that the fixed points $\{|j\rangle\langle j|\}_{j=1}^d$ span a $d$-dimensional subspace. When $\gamma$ is small, $\gamma C$ is a perturbation on $B$, and the $d$-fold degeneration of fixed points breaks down, leaving $r$ fixed points $\{\Pi_j\}_{j=1}^r$ and the other $(d-r)$ metastable points, which corresponds to $n=r$ and $l=d$ in Eq. (\ref{channelmeta}). {\color{black}The EMSs for this case are $ \{|j\rangle\langle j|\}_{j=1}^d$ up to some corrections \cite{Note1}.} Then, metastable polarization occurs when $1/\abs{\ln \abs{\lambda_{d+1}}}\ll m \ll 1/\abs{\ln \abs{\lambda_d}}$, which can also be seen in the measurement statistics.

\textit{Examples.---}First consider that the target system is a single qubit with $B=\sigma_z$ and $C=\sigma_x$, then obviously $[B,C]\neq 0$.
With the target qubit initially in a maximally mixed state $\mathbb{I}/2$, we show the polarization statistics of sequential RIMs by Monte Carlo simulations in Fig. \ref{fig2}. For a small $\gamma$, there appear two distinguishable peaks corresponding to two EMSs as $m$ increases to the meatastable region [Fig. \ref{fig2}{\color{blue}(a-b)}]. While beyond that region, the above two peaks gradually vanish and there appears a single peak corresponding to the stable state $\mathbb{I}/2$ [Fig. \ref{fig2}{\color{blue}(c-d)}]. The polarization plateau in the evolution of $\langle\sigma_z\rangle$ clearly shows the metastability [Fig. \ref{fig2}{\color{blue}(e)}], where the target qubit is polarized to $\ket{0}$ or $\ket{1}$ (eigenstates of $B=\sigma_z$) apart from some corrections. The metastable regions [Fig. \ref{fig2}{\color{blue}(e)}] agree with those predicted by the spectrum of the channel [Fig. \ref{fig2}{\color{blue}(f)}]. Here we divide all the trajectories into three categories by $X\in[-0.5,-0.15)$, $[-0.15,0.15]$ or $(0.15,0.5]$ and obtain the data in Fig. \ref{fig2}{\color{blue}(e)} by averaging all over the trajectories of single categories. EMSs here can be represented as $\rho_{1,2}=\mathbb{I}/2+c_2^{M,m}{R}_2/h$, where $c_2^{M}$ ($c_2^{m}$) is the maximal (minimal) eigenvalue of $L_2$, and $h=\sqrt{\brakett{L_2}{L_2}\brakett{R_2}{R_2}}$ is a normalization coefficient (see \cite{Note1} for details).

Then we consider a target system composed of multiple qubits, such as multiple $^{13}$C nuclear spins coupled to a nitrogen vacancy (NV) center electron spin (as an ancilla) \cite{PhysRevB.85.205203,PhysRevB.85.115303}. In this system, the NV center interacts with $K$ nuclear spins through hyperfine interaction, $B=\sum_{k=1}^K \vb*{A}_k\cdot \vb*{I}_k$, where $\vb*{A}_k=(A_k^{x}, A_k^{y}, A_k^{z})$ and $\vb*{I}_k=(I_k^x,I_k^y,I_k^z)$. The free Hamiltonian of the target system is the dipolar interaction between nuclear spins, $C=\sum_{k<j} D_{jk}\left[\boldsymbol{I}_k \cdot \boldsymbol{I}_j-\frac{3\left(\boldsymbol{I}_k \cdot \boldsymbol{r}_{k j}\right)\left(\boldsymbol{I}_j \cdot \boldsymbol{r}_{k j}\right)}{r_{k j}^2}\right]$ ($\gamma=1$) , where $D_{jk}$ denotes the dipolar coupling strength, and $\vb*{r}_{kj}$ is the displacement from the $i{\rm th}$ to the $j$th target spin. Typically $[B,C]\neq 0$ \cite{PhysRevB.85.115303}, and the only fixed point is the maximally mixed state $\mathbb{I}/2^K$. Since the nuclear dipolar interaction is much smaller than the hyperfine interaction, i.e., $D_{jk}\ll|\vb*{A}_k|$, there will be $2^K-1$ metastable points, forming a $(2^K-1)$-dimensional MM. For measurement statistics of sequential RIMs, there will be $2^K$ peaks if $\mu''\ll m\ll \mu'$, which collapse to a single peak corresponding to $\mathbb{I}/2^K$ as $m\gtrsim\mu'$ (see \cite{Note1} for simulations).

\textit{Discussions.---}Metastability theory still applies when the ancilla is under sequential dynamical decoupling (DD) control \cite{viola1999,ryan2010}. For periodic DD sequences and nearly independent nuclear spins, we have $B=\sum_{k=1}^K A^\perp_k I^\perp_k$, $\quad C=\Delta_\omega\sum_{k=1}^K I_k^z$ ($\gamma=1$), where $A^\perp_k=\sqrt{{A_k^x}^2+{A_k^y}^2}$ and $I_k^\perp=\cos\xi I_k^x+\sin\xi I_k^y$ with $\xi=\arctan(A_k^y/A_k^x)$, and $\Delta_\omega=\omega_L-\omega_T$ denotes the detuning of DD frequency $\omega_L$ relative to the nuclear Larmor frequency $\omega_T$. When $\omega_L$ resonates with $\omega_T$, $C$ can be tuned to zero \cite{ma2016}, leading to the polarization of the nuclear spins \cite{liu2017}. If $\omega_L\neq\omega_T$, the nuclear spins are generally depolarized since $[B,C]\neq 0$, but if $\Delta_\omega$ is small (relative to $A_k$), the nuclear spins can still be polarized for a reasonable range of measurement times (see \cite{Note1} for simulations). Moreover, in all the above examples, numerical simulations show that quantum metastability is quite robust even if the target systems suffers additional dissipations \cite{Note1}.

\textit{Conclusions and outlooks.---}We extend the quantum metastability theory from continuous-time open quantum dynamics described by Lindblad master equations to discrete-time open quantum dynamics described by sequential general quantum channels. We consider the quantum channel induced by both RIM and DD sequences of an ancilla qubit. Metastable polarization behaviors are demonstrated by numerical simulations for a quantum system containing single or multiple qubits. Our work provides theoretical support for quantum state and dynamics engineering with sequential measurement and control of an ancilla system.

In this paper, we focus on the channels generated by RIMs evolving with the Hamiltonian [Eq. (\ref{hamiltonian})], where MM only includes EMSs. It will be interesting future topics to consider more general channels, whose fixed points as a kind of preserved information may include decoherence-free subspaces or noiseless subsystems \cite{Blume-Kohout2008,Blume-Kohout2010}. In the presence of control imperfection or environmental noise, the channel should be slightly perturbed from the ideal ones and the preserved information may become metastable, and we expect that metastability theory in this paper can provide a useful guide to make full use of such metastable preserved information.

We acknowledge support from the National Natural Science Foundation of China (No. 12174379, No. E31Q02BG), the Chinese Academy of Sciences (No. E0SEBB11, No. E27RBB11), the Innovation Program for Quantum Science and Technology (No. 2021ZD0302300) and Chinese Academy of Sciences Project for Young Scientists in Basic Research (YSBR-090).


\appendix

\bibliography{Metastability}

\begin{thebibliography}{87}%
\makeatletter
\providecommand \@ifxundefined [1]{%
 \@ifx{#1\undefined}
}%
\providecommand \@ifnum [1]{%
 \ifnum #1\expandafter \@firstoftwo
 \else \expandafter \@secondoftwo
 \fi
}%
\providecommand \@ifx [1]{%
 \ifx #1\expandafter \@firstoftwo
 \else \expandafter \@secondoftwo
 \fi
}%
\providecommand \natexlab [1]{#1}%
\providecommand \enquote  [1]{``#1''}%
\providecommand \bibnamefont  [1]{#1}%
\providecommand \bibfnamefont [1]{#1}%
\providecommand \citenamefont [1]{#1}%
\providecommand \href@noop [0]{\@secondoftwo}%
\providecommand \href [0]{\begingroup \@sanitize@url \@href}%
\providecommand \@href[1]{\@@startlink{#1}\@@href}%
\providecommand \@@href[1]{\endgroup#1\@@endlink}%
\providecommand \@sanitize@url [0]{\catcode `\\12\catcode `\$12\catcode
  `\&12\catcode `\#12\catcode `\^12\catcode `\_12\catcode `\%12\relax}%
\providecommand \@@startlink[1]{}%
\providecommand \@@endlink[0]{}%
\providecommand \url  [0]{\begingroup\@sanitize@url \@url }%
\providecommand \@url [1]{\endgroup\@href {#1}{\urlprefix }}%
\providecommand \urlprefix  [0]{URL }%
\providecommand \Eprint [0]{\href }%
\providecommand \doibase [0]{https://doi.org/}%
\providecommand \selectlanguage [0]{\@gobble}%
\providecommand \bibinfo  [0]{\@secondoftwo}%
\providecommand \bibfield  [0]{\@secondoftwo}%
\providecommand \translation [1]{[#1]}%
\providecommand \BibitemOpen [0]{}%
\providecommand \bibitemStop [0]{}%
\providecommand \bibitemNoStop [0]{.\EOS\space}%
\providecommand \EOS [0]{\spacefactor3000\relax}%
\providecommand \BibitemShut  [1]{\csname bibitem#1\endcsname}%
\let\auto@bib@innerbib\@empty
\bibitem [{\citenamefont {Gaveau}\ and\ \citenamefont
  {Schulman}(1987)}]{B.Gaveau_1987}%
  \BibitemOpen
  \bibfield  {author} {\bibinfo {author} {\bibfnamefont {B.}~\bibnamefont
  {Gaveau}}\ and\ \bibinfo {author} {\bibfnamefont {L.~S.}\ \bibnamefont
  {Schulman}},\ }\bibfield  {title} {\bibinfo {title} {Dynamical
  metastability},\ }\href {https://doi.org/10.1088/0305-4470/20/10/031}
  {\bibfield  {journal} {\bibinfo  {journal} {J. Phys. A}\ }\textbf {\bibinfo
  {volume} {20}},\ \bibinfo {pages} {2865} (\bibinfo {year}
  {1987})}\BibitemShut {NoStop}%
\bibitem [{\citenamefont {Cheng}\ and\ \citenamefont
  {Keller}(1998)}]{cheng1998}%
  \BibitemOpen
  \bibfield  {author} {\bibinfo {author} {\bibfnamefont {S.~Z.~D.}\
  \bibnamefont {Cheng}}\ and\ \bibinfo {author} {\bibfnamefont
  {A.}~\bibnamefont {Keller}},\ }\bibfield  {title} {\bibinfo {title} {{{The
  role of metastable states in polymer phase transitions}}: {{Concepts}},
  {{Principles}}, and {{Experimental Observations}}},\ }\href
  {https://doi.org/10.1146/annurev.matsci.28.1.533} {\bibfield  {journal}
  {\bibinfo  {journal} {Annu. Rev. Mater. Sci.}\ }\textbf {\bibinfo {volume}
  {28}},\ \bibinfo {pages} {533} (\bibinfo {year} {1998})}\BibitemShut
  {NoStop}%
\bibitem [{\citenamefont {Gaveau}\ and\ \citenamefont
  {Moreau}(2000)}]{Bernard.Gaveau_2000}%
  \BibitemOpen
  \bibfield  {author} {\bibinfo {author} {\bibfnamefont {B.}~\bibnamefont
  {Gaveau}}\ and\ \bibinfo {author} {\bibfnamefont {M.}~\bibnamefont
  {Moreau}},\ }\bibfield  {title} {\bibinfo {title} {Metastable relaxation
  times and absorption probabilities for multidimensional stochastic systems},\
  }\href {https://doi.org/10.1088/0305-4470/33/27/301} {\bibfield  {journal}
  {\bibinfo  {journal} {J. Phys. A}\ }\textbf {\bibinfo {volume} {33}},\
  \bibinfo {pages} {4837} (\bibinfo {year} {2000})}\BibitemShut {NoStop}%
\bibitem [{\citenamefont {Gaveau}\ and\ \citenamefont
  {Schulman}(2006)}]{gaveauMultiplePhasesStochastic2006b}%
  \BibitemOpen
  \bibfield  {author} {\bibinfo {author} {\bibfnamefont {B.}~\bibnamefont
  {Gaveau}}\ and\ \bibinfo {author} {\bibfnamefont {L.~S.}\ \bibnamefont
  {Schulman}},\ }\bibfield  {title} {\bibinfo {title} {Multiple phases in
  stochastic dynamics: {{Geometry}} and probabilities},\ }\href
  {https://doi.org/10.1103/PhysRevE.73.036124} {\bibfield  {journal} {\bibinfo
  {journal} {Phys. Rev. E}\ }\textbf {\bibinfo {volume} {73}},\ \bibinfo
  {pages} {036124} (\bibinfo {year} {2006})}\BibitemShut {NoStop}%
\bibitem [{\citenamefont {Gaveau}\ and\ \citenamefont
  {Schulman}(1998)}]{gaveauTheoryNonequilibriumFirstorder1998a}%
  \BibitemOpen
  \bibfield  {author} {\bibinfo {author} {\bibfnamefont {B.}~\bibnamefont
  {Gaveau}}\ and\ \bibinfo {author} {\bibfnamefont {L.~S.}\ \bibnamefont
  {Schulman}},\ }\bibfield  {title} {\bibinfo {title} {Theory of nonequilibrium
  first-order phase transitions for stochastic dynamics},\ }\href
  {https://doi.org/10.1063/1.532394} {\bibfield  {journal} {\bibinfo  {journal}
  {J. Math. Phys.}\ }\textbf {\bibinfo {volume} {39}},\ \bibinfo {pages} {1517}
  (\bibinfo {year} {1998})}\BibitemShut {NoStop}%
\bibitem [{\citenamefont {Gaveau}\ \emph {et~al.}(1999)\citenamefont {Gaveau},
  \citenamefont {Lesne},\ and\ \citenamefont {Schulman}}]{GAVEAU1999222PT}%
  \BibitemOpen
  \bibfield  {author} {\bibinfo {author} {\bibfnamefont {B.}~\bibnamefont
  {Gaveau}}, \bibinfo {author} {\bibfnamefont {A.}~\bibnamefont {Lesne}},\ and\
  \bibinfo {author} {\bibfnamefont {L.}~\bibnamefont {Schulman}},\ }\bibfield
  {title} {\bibinfo {title} {Spectral signatures of hierarchical relaxation},\
  }\href {https://doi.org/10.1016/S0375-9601(99)00369-2} {\bibfield  {journal}
  {\bibinfo  {journal} {Phys. Lett. A}\ }\textbf {\bibinfo {volume} {258}},\
  \bibinfo {pages} {222} (\bibinfo {year} {1999})}\BibitemShut {NoStop}%
\bibitem [{\citenamefont {Huisinga}\ \emph {et~al.}(2004)\citenamefont
  {Huisinga}, \citenamefont {Meyn},\ and\ \citenamefont
  {Sch{\"u}tte}}]{huisinga2004}%
  \BibitemOpen
  \bibfield  {author} {\bibinfo {author} {\bibfnamefont {W.}~\bibnamefont
  {Huisinga}}, \bibinfo {author} {\bibfnamefont {S.}~\bibnamefont {Meyn}},\
  and\ \bibinfo {author} {\bibfnamefont {C.}~\bibnamefont {Sch{\"u}tte}},\
  }\href@noop {} {\bibfield  {journal} {\bibinfo  {journal} {Ann. Appl.
  Probab.}\ }\textbf {\bibinfo {volume} {14}} (\bibinfo {year}
  {2004})}\BibitemShut {NoStop}%
\bibitem [{\citenamefont {Boettger}\ and\ \citenamefont
  {Wallace}(1997)}]{boettger1997}%
  \BibitemOpen
  \bibfield  {author} {\bibinfo {author} {\bibfnamefont {J.~C.}\ \bibnamefont
  {Boettger}}\ and\ \bibinfo {author} {\bibfnamefont {D.~C.}\ \bibnamefont
  {Wallace}},\ }\bibfield  {title} {\bibinfo {title} {Metastability and
  dynamics of the shock-induced phase transition in iron},\ }\href
  {https://doi.org/10.1103/PhysRevB.55.2840} {\bibfield  {journal} {\bibinfo
  {journal} {Phys. Rev. B}\ }\textbf {\bibinfo {volume} {55}},\ \bibinfo
  {pages} {2840} (\bibinfo {year} {1997})}\BibitemShut {NoStop}%
\bibitem [{\citenamefont {Anderson}(2018)}]{anderson2018}%
  \BibitemOpen
  \bibfield  {author} {\bibinfo {author} {\bibfnamefont {P.~W.}\ \bibnamefont
  {Anderson}},\ }\href {https://doi.org/10.4324/9780429494116} {\emph {\bibinfo
  {title} {Basic {{Notions}} of {{Condensed Matter Physics}}}}},\ \bibinfo
  {edition} {1st}\ ed.,\ edited by\ \bibinfo {editor} {\bibfnamefont {P.~W.}\
  \bibnamefont {Anderson}}\ (\bibinfo  {publisher} {{CRC Press}},\ \bibinfo
  {year} {2018})\BibitemShut {NoStop}%
\bibitem [{\citenamefont {Binder}\ and\ \citenamefont
  {Young}(1986)}]{binder1986}%
  \BibitemOpen
  \bibfield  {author} {\bibinfo {author} {\bibfnamefont {K.}~\bibnamefont
  {Binder}}\ and\ \bibinfo {author} {\bibfnamefont {A.~P.}\ \bibnamefont
  {Young}},\ }\bibfield  {title} {\bibinfo {title} {Spin glasses:
  {{Experimental}} facts, theoretical concepts, and open questions},\ }\href
  {https://doi.org/10.1103/RevModPhys.58.801} {\bibfield  {journal} {\bibinfo
  {journal} {Rev. Mod. Phys.}\ }\textbf {\bibinfo {volume} {58}},\ \bibinfo
  {pages} {801} (\bibinfo {year} {1986})}\BibitemShut {NoStop}%
\bibitem [{\citenamefont {Jckle}\ and\ \citenamefont
  {Eisinger}(1991)}]{jckle1991}%
  \BibitemOpen
  \bibfield  {author} {\bibinfo {author} {\bibfnamefont {J.}~\bibnamefont
  {Jckle}}\ and\ \bibinfo {author} {\bibfnamefont {S.}~\bibnamefont
  {Eisinger}},\ }\bibfield  {title} {\bibinfo {title} {A hierarchically
  constrained kinetic {{Ising}} model},\ }\href
  {https://doi.org/10.1007/BF01453764} {\bibfield  {journal} {\bibinfo
  {journal} {Z. Phys. B: Condens. Matter}\ }\textbf {\bibinfo {volume} {84}},\
  \bibinfo {pages} {115} (\bibinfo {year} {1991})}\BibitemShut {NoStop}%
\bibitem [{\citenamefont {Sollich}\ and\ \citenamefont
  {Evans}(1999)}]{sollich1999}%
  \BibitemOpen
  \bibfield  {author} {\bibinfo {author} {\bibfnamefont {P.}~\bibnamefont
  {Sollich}}\ and\ \bibinfo {author} {\bibfnamefont {M.~R.}\ \bibnamefont
  {Evans}},\ }\bibfield  {title} {\bibinfo {title} {Glassy {{Time-Scale
  Divergence}} and {{Anomalous Coarsening}} in a {{Kinetically Constrained Spin
  Chain}}},\ }\href {https://doi.org/10.1103/PhysRevLett.83.3238} {\bibfield
  {journal} {\bibinfo  {journal} {Phys. Rev. Lett.}\ }\textbf {\bibinfo
  {volume} {83}},\ \bibinfo {pages} {3238} (\bibinfo {year}
  {1999})}\BibitemShut {NoStop}%
\bibitem [{\citenamefont {Cugliandolo}\ and\ \citenamefont
  {Lozano}(1999)}]{cugliandolo1999}%
  \BibitemOpen
  \bibfield  {author} {\bibinfo {author} {\bibfnamefont {L.~F.}\ \bibnamefont
  {Cugliandolo}}\ and\ \bibinfo {author} {\bibfnamefont {G.}~\bibnamefont
  {Lozano}},\ }\bibfield  {title} {\bibinfo {title} {Real-time nonequilibrium
  dynamics of quantum glassy systems},\ }\href
  {https://doi.org/10.1103/PhysRevB.59.915} {\bibfield  {journal} {\bibinfo
  {journal} {Phys. Rev. B}\ }\textbf {\bibinfo {volume} {59}},\ \bibinfo
  {pages} {915} (\bibinfo {year} {1999})}\BibitemShut {NoStop}%
\bibitem [{\citenamefont {Garrahan}\ and\ \citenamefont
  {Chandler}(2002)}]{garrahan2002}%
  \BibitemOpen
  \bibfield  {author} {\bibinfo {author} {\bibfnamefont {J.~P.}\ \bibnamefont
  {Garrahan}}\ and\ \bibinfo {author} {\bibfnamefont {D.}~\bibnamefont
  {Chandler}},\ }\bibfield  {title} {\bibinfo {title} {Geometrical
  {{Explanation}} and {{Scaling}} of {{Dynamical Heterogeneities}} in {{Glass
  Forming Systems}}},\ }\href {https://doi.org/10.1103/PhysRevLett.89.035704}
  {\bibfield  {journal} {\bibinfo  {journal} {Phys. Rev. Lett.}\ }\textbf
  {\bibinfo {volume} {89}},\ \bibinfo {pages} {035704} (\bibinfo {year}
  {2002})}\BibitemShut {NoStop}%
\bibitem [{\citenamefont {Sollich}\ and\ \citenamefont
  {Evans}(2003)}]{sollich2003}%
  \BibitemOpen
  \bibfield  {author} {\bibinfo {author} {\bibfnamefont {P.}~\bibnamefont
  {Sollich}}\ and\ \bibinfo {author} {\bibfnamefont {M.~R.}\ \bibnamefont
  {Evans}},\ }\bibfield  {title} {\bibinfo {title} {Glassy dynamics in the
  asymmetrically constrained kinetic {{Ising}} chain},\ }\href
  {https://doi.org/10.1103/PhysRevE.68.031504} {\bibfield  {journal} {\bibinfo
  {journal} {Phys. Rev. E}\ }\textbf {\bibinfo {volume} {68}},\ \bibinfo
  {pages} {031504} (\bibinfo {year} {2003})}\BibitemShut {NoStop}%
\bibitem [{\citenamefont {Biroli}\ and\ \citenamefont
  {Garrahan}(2013)}]{biroli2013}%
  \BibitemOpen
  \bibfield  {author} {\bibinfo {author} {\bibfnamefont {G.}~\bibnamefont
  {Biroli}}\ and\ \bibinfo {author} {\bibfnamefont {J.~P.}\ \bibnamefont
  {Garrahan}},\ }\bibfield  {title} {\bibinfo {title} {Perspective: {{The}}
  glass transition},\ }\href {https://doi.org/10.1063/1.4795539} {\bibfield
  {journal} {\bibinfo  {journal} {J. Chem. Phys.}\ }\textbf {\bibinfo {volume}
  {138}},\ \bibinfo {pages} {12A301} (\bibinfo {year} {2013})}\BibitemShut
  {NoStop}%
\bibitem [{\citenamefont {Lan}\ \emph {et~al.}(2018)\citenamefont {Lan},
  \citenamefont {Van~Horssen}, \citenamefont {Powell},\ and\ \citenamefont
  {Garrahan}}]{lan2018}%
  \BibitemOpen
  \bibfield  {author} {\bibinfo {author} {\bibfnamefont {Z.}~\bibnamefont
  {Lan}}, \bibinfo {author} {\bibfnamefont {M.}~\bibnamefont {Van~Horssen}},
  \bibinfo {author} {\bibfnamefont {S.}~\bibnamefont {Powell}},\ and\ \bibinfo
  {author} {\bibfnamefont {J.~P.}\ \bibnamefont {Garrahan}},\ }\bibfield
  {title} {\bibinfo {title} {Quantum {{Slow Relaxation}} and {{Metastability}}
  due to {{Dynamical Constraints}}},\ }\href
  {https://doi.org/10.1103/PhysRevLett.121.040603} {\bibfield  {journal}
  {\bibinfo  {journal} {Phys. Rev. Lett.}\ }\textbf {\bibinfo {volume} {121}},\
  \bibinfo {pages} {040603} (\bibinfo {year} {2018})}\BibitemShut {NoStop}%
\bibitem [{\citenamefont {Macieszczak}\ \emph {et~al.}(2016)\citenamefont
  {Macieszczak}, \citenamefont {Gu{\c t}{\u a}}, \citenamefont {Lesanovsky},\
  and\ \citenamefont {Garrahan}}]{PhysRevLett.116.240404}%
  \BibitemOpen
  \bibfield  {author} {\bibinfo {author} {\bibfnamefont {K.}~\bibnamefont
  {Macieszczak}}, \bibinfo {author} {\bibfnamefont {M.}~\bibnamefont {Gu{\c
  t}{\u a}}}, \bibinfo {author} {\bibfnamefont {I.}~\bibnamefont
  {Lesanovsky}},\ and\ \bibinfo {author} {\bibfnamefont {J.~P.}\ \bibnamefont
  {Garrahan}},\ }\bibfield  {title} {\bibinfo {title} {Towards a theory of
  metastability in open quantum dynamics},\ }\href
  {https://doi.org/10.1103/PhysRevLett.116.240404} {\bibfield  {journal}
  {\bibinfo  {journal} {Phys. Rev. Lett.}\ }\textbf {\bibinfo {volume} {116}},\
  \bibinfo {pages} {240404} (\bibinfo {year} {2016})}\BibitemShut {NoStop}%
\bibitem [{\citenamefont {Merkli}\ \emph {et~al.}(2015)\citenamefont {Merkli},
  \citenamefont {Song},\ and\ \citenamefont {Berman}}]{Merkli_2015}%
  \BibitemOpen
  \bibfield  {author} {\bibinfo {author} {\bibfnamefont {M.}~\bibnamefont
  {Merkli}}, \bibinfo {author} {\bibfnamefont {H.}~\bibnamefont {Song}},\ and\
  \bibinfo {author} {\bibfnamefont {G.~P.}\ \bibnamefont {Berman}},\ }\bibfield
   {title} {\bibinfo {title} {Multiscale dynamics of open three-level quantum
  systems with two quasi-degenerate levels},\ }\href
  {https://doi.org/10.1088/1751-8113/48/27/275304} {\bibfield  {journal}
  {\bibinfo  {journal} {J. Phys. A}\ }\textbf {\bibinfo {volume} {48}},\
  \bibinfo {pages} {275304} (\bibinfo {year} {2015})}\BibitemShut {NoStop}%
\bibitem [{\citenamefont {Macieszczak}\ \emph {et~al.}(2021)\citenamefont
  {Macieszczak}, \citenamefont {Rose}, \citenamefont {Lesanovsky},\ and\
  \citenamefont {Garrahan}}]{macieszczak2021}%
  \BibitemOpen
  \bibfield  {author} {\bibinfo {author} {\bibfnamefont {K.}~\bibnamefont
  {Macieszczak}}, \bibinfo {author} {\bibfnamefont {D.~C.}\ \bibnamefont
  {Rose}}, \bibinfo {author} {\bibfnamefont {I.}~\bibnamefont {Lesanovsky}},\
  and\ \bibinfo {author} {\bibfnamefont {J.~P.}\ \bibnamefont {Garrahan}},\
  }\bibfield  {title} {\bibinfo {title} {Theory of classical metastability in
  open quantum systems},\ }\href
  {https://doi.org/10.1103/PhysRevResearch.3.033047} {\bibfield  {journal}
  {\bibinfo  {journal} {Phys. Rev. Research}\ }\textbf {\bibinfo {volume}
  {3}},\ \bibinfo {pages} {033047} (\bibinfo {year} {2021})}\BibitemShut
  {NoStop}%
\bibitem [{\citenamefont {Macieszczak}(2021)}]{macieszczak2021a}%
  \BibitemOpen
  \bibfield  {author} {\bibinfo {author} {\bibfnamefont {K.}~\bibnamefont
  {Macieszczak}},\ }\href@noop {} {\bibinfo {title} {Operational approach to
  metastability}} (\bibinfo {year} {2021}),\ \Eprint
  {https://arxiv.org/abs/2104.05095} {arxiv:2104.05095} \BibitemShut {NoStop}%
\bibitem [{\citenamefont {Brown}\ \emph {et~al.}(2023)\citenamefont {Brown},
  \citenamefont {Macieszczak},\ and\ \citenamefont {Jack}}]{brown2023a}%
  \BibitemOpen
  \bibfield  {author} {\bibinfo {author} {\bibfnamefont {C.~A.}\ \bibnamefont
  {Brown}}, \bibinfo {author} {\bibfnamefont {K.}~\bibnamefont {Macieszczak}},\
  and\ \bibinfo {author} {\bibfnamefont {R.~L.}\ \bibnamefont {Jack}},\
  }\href@noop {} {\bibinfo {title} {Unravelling {{Metastable Markovian Open
  Quantum Systems}}}} (\bibinfo {year} {2023}),\ \Eprint
  {https://arxiv.org/abs/2308.14107} {arxiv:2308.14107 [cond-mat,
  physics:quant-ph]} \BibitemShut {NoStop}%
\bibitem [{\citenamefont {Lindblad}(1976)}]{lindblad1976generators}%
  \BibitemOpen
  \bibfield  {author} {\bibinfo {author} {\bibfnamefont {G.}~\bibnamefont
  {Lindblad}},\ }\bibfield  {title} {\bibinfo {title} {On the generators of
  quantum dynamical semigroups},\ }\href {https://doi.org/10.1007/BF01608499}
  {\bibfield  {journal} {\bibinfo  {journal} {Commun. Math. Phys.}\ }\textbf
  {\bibinfo {volume} {48}},\ \bibinfo {pages} {119} (\bibinfo {year}
  {1976})}\BibitemShut {NoStop}%
\bibitem [{\citenamefont {Gorini}\ \emph {et~al.}(1976)\citenamefont {Gorini},
  \citenamefont {Kossakowski},\ and\ \citenamefont {Sudarshan}}]{gorini1976}%
  \BibitemOpen
  \bibfield  {author} {\bibinfo {author} {\bibfnamefont {V.}~\bibnamefont
  {Gorini}}, \bibinfo {author} {\bibfnamefont {A.}~\bibnamefont
  {Kossakowski}},\ and\ \bibinfo {author} {\bibfnamefont {E.~C.~G.}\
  \bibnamefont {Sudarshan}},\ }\bibfield  {title} {\bibinfo {title} {Completely
  positive dynamical semigroups of {{{\emph{N}}}} -{{Level}} systems},\ }\href
  {https://doi.org/10.1063/1.522979} {\bibfield  {journal} {\bibinfo  {journal}
  {J. Math. Phys.}\ }\textbf {\bibinfo {volume} {17}},\ \bibinfo {pages} {821}
  (\bibinfo {year} {1976})}\BibitemShut {NoStop}%
\bibitem [{\citenamefont {Rose}\ \emph {et~al.}(2016)\citenamefont {Rose},
  \citenamefont {Macieszczak}, \citenamefont {Lesanovsky},\ and\ \citenamefont
  {Garrahan}}]{rose2016}%
  \BibitemOpen
  \bibfield  {author} {\bibinfo {author} {\bibfnamefont {D.~C.}\ \bibnamefont
  {Rose}}, \bibinfo {author} {\bibfnamefont {K.}~\bibnamefont {Macieszczak}},
  \bibinfo {author} {\bibfnamefont {I.}~\bibnamefont {Lesanovsky}},\ and\
  \bibinfo {author} {\bibfnamefont {J.~P.}\ \bibnamefont {Garrahan}},\
  }\bibfield  {title} {\bibinfo {title} {Metastability in an open quantum
  {{Ising}} model},\ }\href {https://doi.org/10.1103/PhysRevE.94.052132}
  {\bibfield  {journal} {\bibinfo  {journal} {Phys. Rev. E}\ }\textbf {\bibinfo
  {volume} {94}},\ \bibinfo {pages} {052132} (\bibinfo {year}
  {2016})}\BibitemShut {NoStop}%
\bibitem [{\citenamefont {J{\"a}ger}\ \emph {et~al.}(2022)\citenamefont
  {J{\"a}ger}, \citenamefont {Schmit}, \citenamefont {Morigi}, \citenamefont
  {Holland},\ and\ \citenamefont {Betzholz}}]{PhysRevLett.129.063601}%
  \BibitemOpen
  \bibfield  {author} {\bibinfo {author} {\bibfnamefont {S.~B.}\ \bibnamefont
  {J{\"a}ger}}, \bibinfo {author} {\bibfnamefont {T.}~\bibnamefont {Schmit}},
  \bibinfo {author} {\bibfnamefont {G.}~\bibnamefont {Morigi}}, \bibinfo
  {author} {\bibfnamefont {M.~J.}\ \bibnamefont {Holland}},\ and\ \bibinfo
  {author} {\bibfnamefont {R.}~\bibnamefont {Betzholz}},\ }\bibfield  {title}
  {\bibinfo {title} {Lindblad master equations for quantum systems coupled to
  dissipative bosonic modes},\ }\href
  {https://doi.org/10.1103/PhysRevLett.129.063601} {\bibfield  {journal}
  {\bibinfo  {journal} {Phys. Rev. Lett.}\ }\textbf {\bibinfo {volume} {129}},\
  \bibinfo {pages} {063601} (\bibinfo {year} {2022})}\BibitemShut {NoStop}%
\bibitem [{\citenamefont {Le~Boit{\'e}}\ \emph {et~al.}(2017)\citenamefont
  {Le~Boit{\'e}}, \citenamefont {Hwang},\ and\ \citenamefont
  {Plenio}}]{leboite2017}%
  \BibitemOpen
  \bibfield  {author} {\bibinfo {author} {\bibfnamefont {A.}~\bibnamefont
  {Le~Boit{\'e}}}, \bibinfo {author} {\bibfnamefont {M.-J.}\ \bibnamefont
  {Hwang}},\ and\ \bibinfo {author} {\bibfnamefont {M.~B.}\ \bibnamefont
  {Plenio}},\ }\bibfield  {title} {\bibinfo {title} {Metastability in the
  driven-dissipative {{Rabi}} model},\ }\href
  {https://doi.org/10.1103/PhysRevA.95.023829} {\bibfield  {journal} {\bibinfo
  {journal} {Phys. Rev. A}\ }\textbf {\bibinfo {volume} {95}},\ \bibinfo
  {pages} {023829} (\bibinfo {year} {2017})}\BibitemShut {NoStop}%
\bibitem [{\citenamefont {Liu}\ \emph {et~al.}(2022)\citenamefont {Liu},
  \citenamefont {Zhang}, \citenamefont {Xu}, \citenamefont {Cui},\ and\
  \citenamefont {Fan}}]{liu2022}%
  \BibitemOpen
  \bibfield  {author} {\bibinfo {author} {\bibfnamefont {T.}~\bibnamefont
  {Liu}}, \bibinfo {author} {\bibfnamefont {Y.-R.}\ \bibnamefont {Zhang}},
  \bibinfo {author} {\bibfnamefont {K.}~\bibnamefont {Xu}}, \bibinfo {author}
  {\bibfnamefont {J.}~\bibnamefont {Cui}},\ and\ \bibinfo {author}
  {\bibfnamefont {H.}~\bibnamefont {Fan}},\ }\bibfield  {title} {\bibinfo
  {title} {Discrete time crystal in a driven-dissipative {{Bose-Hubbard}} model
  with two-photon processes},\ }\href
  {https://doi.org/10.1103/PhysRevA.105.013710} {\bibfield  {journal} {\bibinfo
   {journal} {Phys. Rev. A}\ }\textbf {\bibinfo {volume} {105}},\ \bibinfo
  {pages} {013710} (\bibinfo {year} {2022})}\BibitemShut {NoStop}%
\bibitem [{\citenamefont {{Nicol{\`o}
  Defenu}}(2021)}]{doi:10.1073/pnas.2101785118}%
  \BibitemOpen
  \bibfield  {author} {\bibinfo {author} {\bibnamefont {{Nicol{\`o} Defenu}}},\
  }\bibfield  {title} {\bibinfo {title} {Metastability and discrete spectrum of
  long-range systems},\ }\href {https://doi.org/10.1073/pnas.2101785118}
  {\bibfield  {journal} {\bibinfo  {journal} {Proc. Natl. Acad. Sci.}\ }\textbf
  {\bibinfo {volume} {118}},\ \bibinfo {pages} {e2101785118} (\bibinfo {year}
  {2021})}\BibitemShut {NoStop}%
\bibitem [{\citenamefont {Letscher}\ \emph {et~al.}(2017)\citenamefont
  {Letscher}, \citenamefont {Thomas}, \citenamefont {Niederpr{\"u}m},
  \citenamefont {Fleischhauer},\ and\ \citenamefont {Ott}}]{letscher2017}%
  \BibitemOpen
  \bibfield  {author} {\bibinfo {author} {\bibfnamefont {F.}~\bibnamefont
  {Letscher}}, \bibinfo {author} {\bibfnamefont {O.}~\bibnamefont {Thomas}},
  \bibinfo {author} {\bibfnamefont {T.}~\bibnamefont {Niederpr{\"u}m}},
  \bibinfo {author} {\bibfnamefont {M.}~\bibnamefont {Fleischhauer}},\ and\
  \bibinfo {author} {\bibfnamefont {H.}~\bibnamefont {Ott}},\ }\bibfield
  {title} {\bibinfo {title} {Bistability {{Versus Metastability}} in {{Driven
  Dissipative Rydberg Gases}}},\ }\href
  {https://doi.org/10.1103/PhysRevX.7.021020} {\bibfield  {journal} {\bibinfo
  {journal} {Phys. Rev. X}\ }\textbf {\bibinfo {volume} {7}},\ \bibinfo {pages}
  {021020} (\bibinfo {year} {2017})}\BibitemShut {NoStop}%
\bibitem [{\citenamefont {Gong}\ \emph {et~al.}(2018)\citenamefont {Gong},
  \citenamefont {Hamazaki},\ and\ \citenamefont {Ueda}}]{Gong2018}%
  \BibitemOpen
  \bibfield  {author} {\bibinfo {author} {\bibfnamefont {Z.}~\bibnamefont
  {Gong}}, \bibinfo {author} {\bibfnamefont {R.}~\bibnamefont {Hamazaki}},\
  and\ \bibinfo {author} {\bibfnamefont {M.}~\bibnamefont {Ueda}},\ }\bibfield
  {title} {\bibinfo {title} {Discrete time-crystalline order in cavity and
  circuit qed systems},\ }\href
  {https://doi.org/10.1103/PhysRevLett.120.040404} {\bibfield  {journal}
  {\bibinfo  {journal} {Phys. Rev. Lett.}\ }\textbf {\bibinfo {volume} {120}},\
  \bibinfo {pages} {040404} (\bibinfo {year} {2018})}\BibitemShut {NoStop}%
\bibitem [{\citenamefont {Gambetta}\ \emph {et~al.}(2019)\citenamefont
  {Gambetta}, \citenamefont {Carollo}, \citenamefont {Marcuzzi}, \citenamefont
  {Garrahan},\ and\ \citenamefont {Lesanovsky}}]{Gambetta2019}%
  \BibitemOpen
  \bibfield  {author} {\bibinfo {author} {\bibfnamefont {F.~M.}\ \bibnamefont
  {Gambetta}}, \bibinfo {author} {\bibfnamefont {F.}~\bibnamefont {Carollo}},
  \bibinfo {author} {\bibfnamefont {M.}~\bibnamefont {Marcuzzi}}, \bibinfo
  {author} {\bibfnamefont {J.~P.}\ \bibnamefont {Garrahan}},\ and\ \bibinfo
  {author} {\bibfnamefont {I.}~\bibnamefont {Lesanovsky}},\ }\bibfield  {title}
  {\bibinfo {title} {Discrete time crystals in the absence of manifest
  symmetries or disorder in open quantum systems},\ }\href
  {https://doi.org/10.1103/PhysRevLett.122.015701} {\bibfield  {journal}
  {\bibinfo  {journal} {Phys. Rev. Lett.}\ }\textbf {\bibinfo {volume} {122}},\
  \bibinfo {pages} {015701} (\bibinfo {year} {2019})}\BibitemShut {NoStop}%
\bibitem [{\citenamefont {Cabot}\ \emph {et~al.}(2022)\citenamefont {Cabot},
  \citenamefont {Carollo},\ and\ \citenamefont {Lesanovsky}}]{cabot2022}%
  \BibitemOpen
  \bibfield  {author} {\bibinfo {author} {\bibfnamefont {A.}~\bibnamefont
  {Cabot}}, \bibinfo {author} {\bibfnamefont {F.}~\bibnamefont {Carollo}},\
  and\ \bibinfo {author} {\bibfnamefont {I.}~\bibnamefont {Lesanovsky}},\
  }\bibfield  {title} {\bibinfo {title} {Metastable discrete time-crystal
  resonances in a dissipative central spin system},\ }\href
  {https://doi.org/10.1103/PhysRevB.106.134311} {\bibfield  {journal} {\bibinfo
   {journal} {Phys. Rev. B}\ }\textbf {\bibinfo {volume} {106}},\ \bibinfo
  {pages} {134311} (\bibinfo {year} {2022})}\BibitemShut {NoStop}%
\bibitem [{\citenamefont {Flynn}\ \emph {et~al.}(2021)\citenamefont {Flynn},
  \citenamefont {Cobanera},\ and\ \citenamefont {Viola}}]{flynn2021}%
  \BibitemOpen
  \bibfield  {author} {\bibinfo {author} {\bibfnamefont {V.~P.}\ \bibnamefont
  {Flynn}}, \bibinfo {author} {\bibfnamefont {E.}~\bibnamefont {Cobanera}},\
  and\ \bibinfo {author} {\bibfnamefont {L.}~\bibnamefont {Viola}},\ }\bibfield
   {title} {\bibinfo {title} {Topology by {{Dissipation}}: {{Majorana Bosons}}
  in {{Metastable Quadratic Markovian Dynamics}}},\ }\href
  {https://doi.org/10.1103/PhysRevLett.127.245701} {\bibfield  {journal}
  {\bibinfo  {journal} {Phys. Rev. Lett.}\ }\textbf {\bibinfo {volume} {127}},\
  \bibinfo {pages} {245701} (\bibinfo {year} {2021})}\BibitemShut {NoStop}%
\bibitem [{\citenamefont {Gudder}(2008)}]{gudder2008}%
  \BibitemOpen
  \bibfield  {author} {\bibinfo {author} {\bibfnamefont {S.}~\bibnamefont
  {Gudder}},\ }\bibfield  {title} {\bibinfo {title} {Quantum {{Markov}}
  chains},\ }\href {https://doi.org/10.1063/1.2953952} {\bibfield  {journal}
  {\bibinfo  {journal} {J. Math. Phys.}\ }\textbf {\bibinfo {volume} {49}},\
  \bibinfo {pages} {072105} (\bibinfo {year} {2008})}\BibitemShut {NoStop}%
\bibitem [{\citenamefont {Guan}\ \emph {et~al.}(2018)\citenamefont {Guan},
  \citenamefont {Feng},\ and\ \citenamefont {Ying}}]{guan2018}%
  \BibitemOpen
  \bibfield  {author} {\bibinfo {author} {\bibfnamefont {J.}~\bibnamefont
  {Guan}}, \bibinfo {author} {\bibfnamefont {Y.}~\bibnamefont {Feng}},\ and\
  \bibinfo {author} {\bibfnamefont {M.}~\bibnamefont {Ying}},\ }\bibfield
  {title} {\bibinfo {title} {Decomposition of quantum {{Markov}} chains and its
  applications},\ }\href {https://doi.org/10.1016/j.jcss.2018.01.005}
  {\bibfield  {journal} {\bibinfo  {journal} {J. Comput. Syst. Sci.}\ }\textbf
  {\bibinfo {volume} {95}},\ \bibinfo {pages} {55} (\bibinfo {year}
  {2018})}\BibitemShut {NoStop}%
\bibitem [{\citenamefont {Novotn{\'{y}}}\ \emph {et~al.}(2018)\citenamefont
  {Novotn{\'{y}}}, \citenamefont {Mary\v{s}ka},\ and\ \citenamefont
  {Jex}}]{novotny2018quantum}%
  \BibitemOpen
  \bibfield  {author} {\bibinfo {author} {\bibfnamefont {J.}~\bibnamefont
  {Novotn{\'{y}}}}, \bibinfo {author} {\bibfnamefont {J.}~\bibnamefont
  {Mary\v{s}ka}},\ and\ \bibinfo {author} {\bibfnamefont {I.}~\bibnamefont
  {Jex}},\ }\bibfield  {title} {\bibinfo {title} {Quantum markov processes:
  From attractor structure to explicit forms of asymptotic states: Asymptotic
  dynamics of quantum markov processes},\ }\href
  {https://doi.org/10.1140/epjp/i2018-12109-8} {\bibfield  {journal} {\bibinfo
  {journal} {Eur. Phys. J. Plus}\ }\textbf {\bibinfo {volume} {133}},\ \bibinfo
  {pages} {1} (\bibinfo {year} {2018})}\BibitemShut {NoStop}%
\bibitem [{\citenamefont {Amato}\ \emph {et~al.}(2023)\citenamefont {Amato},
  \citenamefont {Facchi},\ and\ \citenamefont {Konderak}}]{amato2023}%
  \BibitemOpen
  \bibfield  {author} {\bibinfo {author} {\bibfnamefont {D.}~\bibnamefont
  {Amato}}, \bibinfo {author} {\bibfnamefont {P.}~\bibnamefont {Facchi}},\ and\
  \bibinfo {author} {\bibfnamefont {A.}~\bibnamefont {Konderak}},\ }\bibfield
  {title} {\bibinfo {title} {Asymptotics of quantum channels},\ }\href
  {https://doi.org/10.1088/1751-8121/acd828} {\bibfield  {journal} {\bibinfo
  {journal} {J. Phys. A: Math. Theor.}\ }\textbf {\bibinfo {volume} {56}},\
  \bibinfo {pages} {265304} (\bibinfo {year} {2023})}\BibitemShut {NoStop}%
\bibitem [{\citenamefont {Attal}\ \emph {et~al.}(2012)\citenamefont {Attal},
  \citenamefont {Petruccione}, \citenamefont {Sabot},\ and\ \citenamefont
  {Sinayskiy}}]{attal2012}%
  \BibitemOpen
  \bibfield  {author} {\bibinfo {author} {\bibfnamefont {S.}~\bibnamefont
  {Attal}}, \bibinfo {author} {\bibfnamefont {F.}~\bibnamefont {Petruccione}},
  \bibinfo {author} {\bibfnamefont {C.}~\bibnamefont {Sabot}},\ and\ \bibinfo
  {author} {\bibfnamefont {I.}~\bibnamefont {Sinayskiy}},\ }\bibfield  {title}
  {\bibinfo {title} {Open {{Quantum Random Walks}}},\ }\href
  {https://doi.org/10.1007/s10955-012-0491-0} {\bibfield  {journal} {\bibinfo
  {journal} {J. Stat. Phys.}\ }\textbf {\bibinfo {volume} {147}},\ \bibinfo
  {pages} {832} (\bibinfo {year} {2012})}\BibitemShut {NoStop}%
\bibitem [{\citenamefont {Ciccarello}\ \emph {et~al.}(2022)\citenamefont
  {Ciccarello}, \citenamefont {Lorenzo}, \citenamefont {Giovannetti},\ and\
  \citenamefont {Palma}}]{ciccarello2022}%
  \BibitemOpen
  \bibfield  {author} {\bibinfo {author} {\bibfnamefont {F.}~\bibnamefont
  {Ciccarello}}, \bibinfo {author} {\bibfnamefont {S.}~\bibnamefont {Lorenzo}},
  \bibinfo {author} {\bibfnamefont {V.}~\bibnamefont {Giovannetti}},\ and\
  \bibinfo {author} {\bibfnamefont {G.~M.}\ \bibnamefont {Palma}},\ }\bibfield
  {title} {\bibinfo {title} {Quantum collision models: {{Open}} system dynamics
  from repeated interactions},\ }\href
  {https://doi.org/10.1016/j.physrep.2022.01.001} {\bibfield  {journal}
  {\bibinfo  {journal} {Phys. Rep.}\ }\textbf {\bibinfo {volume} {954}},\
  \bibinfo {pages} {1} (\bibinfo {year} {2022})}\BibitemShut {NoStop}%
\bibitem [{\citenamefont {Lloyd}\ and\ \citenamefont
  {Viola}(2001)}]{lloyd2001}%
  \BibitemOpen
  \bibfield  {author} {\bibinfo {author} {\bibfnamefont {S.}~\bibnamefont
  {Lloyd}}\ and\ \bibinfo {author} {\bibfnamefont {L.}~\bibnamefont {Viola}},\
  }\bibfield  {title} {\bibinfo {title} {Engineering quantum dynamics},\ }\href
  {https://doi.org/10.1103/PhysRevA.65.010101} {\bibfield  {journal} {\bibinfo
  {journal} {Phys. Rev. A}\ }\textbf {\bibinfo {volume} {65}},\ \bibinfo
  {pages} {010101} (\bibinfo {year} {2001})}\BibitemShut {NoStop}%
\bibitem [{\citenamefont {Andersson}\ and\ \citenamefont
  {Oi}(2008)}]{andersson2008}%
  \BibitemOpen
  \bibfield  {author} {\bibinfo {author} {\bibfnamefont {E.}~\bibnamefont
  {Andersson}}\ and\ \bibinfo {author} {\bibfnamefont {D.~K.~L.}\ \bibnamefont
  {Oi}},\ }\bibfield  {title} {\bibinfo {title} {Binary search trees for
  generalized measurements},\ }\href
  {https://doi.org/10.1103/PhysRevA.77.052104} {\bibfield  {journal} {\bibinfo
  {journal} {Phys. Rev. A}\ }\textbf {\bibinfo {volume} {77}},\ \bibinfo
  {pages} {052104} (\bibinfo {year} {2008})}\BibitemShut {NoStop}%
\bibitem [{\citenamefont {Shen}\ \emph {et~al.}(2017)\citenamefont {Shen},
  \citenamefont {Noh}, \citenamefont {Albert}, \citenamefont {Krastanov},
  \citenamefont {Devoret}, \citenamefont {Schoelkopf}, \citenamefont {Girvin},\
  and\ \citenamefont {Jiang}}]{shen2017}%
  \BibitemOpen
  \bibfield  {author} {\bibinfo {author} {\bibfnamefont {C.}~\bibnamefont
  {Shen}}, \bibinfo {author} {\bibfnamefont {K.}~\bibnamefont {Noh}}, \bibinfo
  {author} {\bibfnamefont {V.~V.}\ \bibnamefont {Albert}}, \bibinfo {author}
  {\bibfnamefont {S.}~\bibnamefont {Krastanov}}, \bibinfo {author}
  {\bibfnamefont {M.~H.}\ \bibnamefont {Devoret}}, \bibinfo {author}
  {\bibfnamefont {R.~J.}\ \bibnamefont {Schoelkopf}}, \bibinfo {author}
  {\bibfnamefont {S.~M.}\ \bibnamefont {Girvin}},\ and\ \bibinfo {author}
  {\bibfnamefont {L.}~\bibnamefont {Jiang}},\ }\bibfield  {title} {\bibinfo
  {title} {Quantum channel construction with circuit quantum electrodynamics},\
  }\href {https://doi.org/10.1103/PhysRevB.95.134501} {\bibfield  {journal}
  {\bibinfo  {journal} {Phys. Rev. B}\ }\textbf {\bibinfo {volume} {95}},\
  \bibinfo {pages} {134501} (\bibinfo {year} {2017})}\BibitemShut {NoStop}%
\bibitem [{\citenamefont {Han}\ \emph {et~al.}(2021)\citenamefont {Han},
  \citenamefont {Cai}, \citenamefont {Hu}, \citenamefont {Mu}, \citenamefont
  {Ma}, \citenamefont {Xu}, \citenamefont {Wang}, \citenamefont {Wang},
  \citenamefont {Song}, \citenamefont {Zou},\ and\ \citenamefont
  {Sun}}]{han2021}%
  \BibitemOpen
  \bibfield  {author} {\bibinfo {author} {\bibfnamefont {J.}~\bibnamefont
  {Han}}, \bibinfo {author} {\bibfnamefont {W.}~\bibnamefont {Cai}}, \bibinfo
  {author} {\bibfnamefont {L.}~\bibnamefont {Hu}}, \bibinfo {author}
  {\bibfnamefont {X.}~\bibnamefont {Mu}}, \bibinfo {author} {\bibfnamefont
  {Y.}~\bibnamefont {Ma}}, \bibinfo {author} {\bibfnamefont {Y.}~\bibnamefont
  {Xu}}, \bibinfo {author} {\bibfnamefont {W.}~\bibnamefont {Wang}}, \bibinfo
  {author} {\bibfnamefont {H.}~\bibnamefont {Wang}}, \bibinfo {author}
  {\bibfnamefont {Y.~P.}\ \bibnamefont {Song}}, \bibinfo {author}
  {\bibfnamefont {C.-L.}\ \bibnamefont {Zou}},\ and\ \bibinfo {author}
  {\bibfnamefont {L.}~\bibnamefont {Sun}},\ }\bibfield  {title} {\bibinfo
  {title} {Experimental {{Simulation}} of {{Open Quantum System Dynamics}} via
  {{Trotterization}}},\ }\href {https://doi.org/10.1103/PhysRevLett.127.020504}
  {\bibfield  {journal} {\bibinfo  {journal} {Phys. Rev. Lett.}\ }\textbf
  {\bibinfo {volume} {127}},\ \bibinfo {pages} {020504} (\bibinfo {year}
  {2021})}\BibitemShut {NoStop}%
\bibitem [{\citenamefont {Cai}\ \emph {et~al.}(2021)\citenamefont {Cai},
  \citenamefont {Han}, \citenamefont {Hu}, \citenamefont {Ma}, \citenamefont
  {Mu}, \citenamefont {Wang}, \citenamefont {Xu}, \citenamefont {Hua},
  \citenamefont {Wang}, \citenamefont {Song}, \citenamefont {Zhang},
  \citenamefont {Zou},\ and\ \citenamefont {Sun}}]{cai2021}%
  \BibitemOpen
  \bibfield  {author} {\bibinfo {author} {\bibfnamefont {W.}~\bibnamefont
  {Cai}}, \bibinfo {author} {\bibfnamefont {J.}~\bibnamefont {Han}}, \bibinfo
  {author} {\bibfnamefont {L.}~\bibnamefont {Hu}}, \bibinfo {author}
  {\bibfnamefont {Y.}~\bibnamefont {Ma}}, \bibinfo {author} {\bibfnamefont
  {X.}~\bibnamefont {Mu}}, \bibinfo {author} {\bibfnamefont {W.}~\bibnamefont
  {Wang}}, \bibinfo {author} {\bibfnamefont {Y.}~\bibnamefont {Xu}}, \bibinfo
  {author} {\bibfnamefont {Z.}~\bibnamefont {Hua}}, \bibinfo {author}
  {\bibfnamefont {H.}~\bibnamefont {Wang}}, \bibinfo {author} {\bibfnamefont
  {Y.~P.}\ \bibnamefont {Song}}, \bibinfo {author} {\bibfnamefont {J.-N.}\
  \bibnamefont {Zhang}}, \bibinfo {author} {\bibfnamefont {C.-L.}\ \bibnamefont
  {Zou}},\ and\ \bibinfo {author} {\bibfnamefont {L.}~\bibnamefont {Sun}},\
  }\bibfield  {title} {\bibinfo {title} {High-{{Efficiency Arbitrary Quantum
  Operation}} on a {{High-Dimensional Quantum System}}},\ }\href
  {https://doi.org/10.1103/PhysRevLett.127.090504} {\bibfield  {journal}
  {\bibinfo  {journal} {Phys. Rev. Lett.}\ }\textbf {\bibinfo {volume} {127}},\
  \bibinfo {pages} {090504} (\bibinfo {year} {2021})}\BibitemShut {NoStop}%
\bibitem [{\citenamefont {Degen}\ \emph {et~al.}(2017)\citenamefont {Degen},
  \citenamefont {Reinhard},\ and\ \citenamefont {Cappellaro}}]{degen2017}%
  \BibitemOpen
  \bibfield  {author} {\bibinfo {author} {\bibfnamefont {C.~L.}\ \bibnamefont
  {Degen}}, \bibinfo {author} {\bibfnamefont {F.}~\bibnamefont {Reinhard}},\
  and\ \bibinfo {author} {\bibfnamefont {P.}~\bibnamefont {Cappellaro}},\
  }\bibfield  {title} {\bibinfo {title} {Quantum sensing},\ }\href
  {https://doi.org/10.1103/RevModPhys.89.035002} {\bibfield  {journal}
  {\bibinfo  {journal} {Rev. Mod. Phys.}\ }\textbf {\bibinfo {volume} {89}},\
  \bibinfo {pages} {035002} (\bibinfo {year} {2017})}\BibitemShut {NoStop}%
\bibitem [{\citenamefont {Dasari}\ \emph {et~al.}(2022)\citenamefont {Dasari},
  \citenamefont {Yang}, \citenamefont {Chakrabarti}, \citenamefont {Finkler},
  \citenamefont {Kurizki},\ and\ \citenamefont {Wrachtrup}}]{dasari2022}%
  \BibitemOpen
  \bibfield  {author} {\bibinfo {author} {\bibfnamefont {D.~B.~R.}\
  \bibnamefont {Dasari}}, \bibinfo {author} {\bibfnamefont {S.}~\bibnamefont
  {Yang}}, \bibinfo {author} {\bibfnamefont {A.}~\bibnamefont {Chakrabarti}},
  \bibinfo {author} {\bibfnamefont {A.}~\bibnamefont {Finkler}}, \bibinfo
  {author} {\bibfnamefont {G.}~\bibnamefont {Kurizki}},\ and\ \bibinfo {author}
  {\bibfnamefont {J.}~\bibnamefont {Wrachtrup}},\ }\bibfield  {title} {\bibinfo
  {title} {Anti-{{Zeno}} purification of spin baths by quantum probe
  measurements},\ }\href {https://doi.org/10.1038/s41467-022-35045-3}
  {\bibfield  {journal} {\bibinfo  {journal} {Nat. Commun.}\ }\textbf {\bibinfo
  {volume} {13}},\ \bibinfo {pages} {7527} (\bibinfo {year}
  {2022})}\BibitemShut {NoStop}%
\bibitem [{\citenamefont {M{\k{a}}dzik}\ \emph {et~al.}(2020)\citenamefont
  {M{\k{a}}dzik}, \citenamefont {Ladd}, \citenamefont {Hudson}, \citenamefont
  {Itoh}, \citenamefont {Jakob}, \citenamefont {Johnson}, \citenamefont
  {McCallum}, \citenamefont {Jamieson}, \citenamefont {Dzurak}, \citenamefont
  {Laucht},\ and\ \citenamefont {Morello}}]{madzik2020}%
  \BibitemOpen
  \bibfield  {author} {\bibinfo {author} {\bibfnamefont {M.~T.}\ \bibnamefont
  {M{\k{a}}dzik}}, \bibinfo {author} {\bibfnamefont {T.~D.}\ \bibnamefont
  {Ladd}}, \bibinfo {author} {\bibfnamefont {F.~E.}\ \bibnamefont {Hudson}},
  \bibinfo {author} {\bibfnamefont {K.~M.}\ \bibnamefont {Itoh}}, \bibinfo
  {author} {\bibfnamefont {A.~M.}\ \bibnamefont {Jakob}}, \bibinfo {author}
  {\bibfnamefont {B.~C.}\ \bibnamefont {Johnson}}, \bibinfo {author}
  {\bibfnamefont {J.~C.}\ \bibnamefont {McCallum}}, \bibinfo {author}
  {\bibfnamefont {D.~N.}\ \bibnamefont {Jamieson}}, \bibinfo {author}
  {\bibfnamefont {A.~S.}\ \bibnamefont {Dzurak}}, \bibinfo {author}
  {\bibfnamefont {A.}~\bibnamefont {Laucht}},\ and\ \bibinfo {author}
  {\bibfnamefont {A.}~\bibnamefont {Morello}},\ }\bibfield  {title} {\bibinfo
  {title} {Controllable freezing of the nuclear spin bath in a single-atom spin
  qubit},\ }\href {https://doi.org/10.1126/sciadv.aba3442} {\bibfield
  {journal} {\bibinfo  {journal} {Sci. Adv.}\ }\textbf {\bibinfo {volume}
  {6}},\ \bibinfo {pages} {eaba3442} (\bibinfo {year} {2020})}\BibitemShut
  {NoStop}%
\bibitem [{\citenamefont {Liu}\ \emph {et~al.}(2017)\citenamefont {Liu},
  \citenamefont {Xing}, \citenamefont {Ma}, \citenamefont {Wang}, \citenamefont
  {Li}, \citenamefont {Po}, \citenamefont {Zhang}, \citenamefont {Fan},
  \citenamefont {Liu},\ and\ \citenamefont {Pan}}]{liu2017}%
  \BibitemOpen
  \bibfield  {author} {\bibinfo {author} {\bibfnamefont {G.-Q.}\ \bibnamefont
  {Liu}}, \bibinfo {author} {\bibfnamefont {J.}~\bibnamefont {Xing}}, \bibinfo
  {author} {\bibfnamefont {W.-L.}\ \bibnamefont {Ma}}, \bibinfo {author}
  {\bibfnamefont {P.}~\bibnamefont {Wang}}, \bibinfo {author} {\bibfnamefont
  {C.-H.}\ \bibnamefont {Li}}, \bibinfo {author} {\bibfnamefont {H.~C.}\
  \bibnamefont {Po}}, \bibinfo {author} {\bibfnamefont {Y.-R.}\ \bibnamefont
  {Zhang}}, \bibinfo {author} {\bibfnamefont {H.}~\bibnamefont {Fan}}, \bibinfo
  {author} {\bibfnamefont {R.-B.}\ \bibnamefont {Liu}},\ and\ \bibinfo {author}
  {\bibfnamefont {X.-Y.}\ \bibnamefont {Pan}},\ }\bibfield  {title} {\bibinfo
  {title} {Single-{{Shot Readout}} of a {{Nuclear Spin Weakly Coupled}} to a
  {{Nitrogen-Vacancy Center}} at {{Room Temperature}}},\ }\href
  {https://doi.org/10.1103/PhysRevLett.118.150504} {\bibfield  {journal}
  {\bibinfo  {journal} {Phys. Rev. Lett.}\ }\textbf {\bibinfo {volume} {118}},\
  \bibinfo {pages} {150504} (\bibinfo {year} {2017})}\BibitemShut {NoStop}%
\bibitem [{\citenamefont {Bhaktavatsala~Rao}\ \emph {et~al.}(2019)\citenamefont
  {Bhaktavatsala~Rao}, \citenamefont {Yang}, \citenamefont {Jesenski},
  \citenamefont {Tekin}, \citenamefont {Kaiser},\ and\ \citenamefont
  {Wrachtrup}}]{bhaktavatsalarao2019}%
  \BibitemOpen
  \bibfield  {author} {\bibinfo {author} {\bibfnamefont {D.~D.}\ \bibnamefont
  {Bhaktavatsala~Rao}}, \bibinfo {author} {\bibfnamefont {S.}~\bibnamefont
  {Yang}}, \bibinfo {author} {\bibfnamefont {S.}~\bibnamefont {Jesenski}},
  \bibinfo {author} {\bibfnamefont {E.}~\bibnamefont {Tekin}}, \bibinfo
  {author} {\bibfnamefont {F.}~\bibnamefont {Kaiser}},\ and\ \bibinfo {author}
  {\bibfnamefont {J.}~\bibnamefont {Wrachtrup}},\ }\bibfield  {title} {\bibinfo
  {title} {Observation of nonclassical measurement statistics induced by a
  coherent spin environment},\ }\href
  {https://doi.org/10.1103/PhysRevA.100.022307} {\bibfield  {journal} {\bibinfo
   {journal} {Phys. Rev. A}\ }\textbf {\bibinfo {volume} {100}},\ \bibinfo
  {pages} {022307} (\bibinfo {year} {2019})}\BibitemShut {NoStop}%
\bibitem [{\citenamefont {Kraus}\ \emph {et~al.}(1983)\citenamefont {Kraus},
  \citenamefont {B{\"o}hm}, \citenamefont {Dollard},\ and\ \citenamefont
  {Wootters}}]{kraus1983states}%
  \BibitemOpen
  \bibfield  {author} {\bibinfo {author} {\bibfnamefont {K.}~\bibnamefont
  {Kraus}}, \bibinfo {author} {\bibfnamefont {A.}~\bibnamefont {B{\"o}hm}},
  \bibinfo {author} {\bibfnamefont {J.~D.}\ \bibnamefont {Dollard}},\ and\
  \bibinfo {author} {\bibfnamefont {{\relax WH}.}~\bibnamefont {Wootters}},\
  }\href@noop {} {\emph {\bibinfo {title} {States, Effects, and Operations
  Fundamental Notions of Quantum Theory: {{Lectures}} in Mathematical Physics
  at the University of Texas at Austin}}}\ (\bibinfo  {publisher}
  {{Springer}},\ \bibinfo {year} {1983})\BibitemShut {NoStop}%
\bibitem [{\citenamefont {Caruso}\ \emph {et~al.}(2014)\citenamefont {Caruso},
  \citenamefont {Giovannetti}, \citenamefont {Lupo},\ and\ \citenamefont
  {Mancini}}]{caruso2014quantum}%
  \BibitemOpen
  \bibfield  {author} {\bibinfo {author} {\bibfnamefont {F.}~\bibnamefont
  {Caruso}}, \bibinfo {author} {\bibfnamefont {V.}~\bibnamefont {Giovannetti}},
  \bibinfo {author} {\bibfnamefont {C.}~\bibnamefont {Lupo}},\ and\ \bibinfo
  {author} {\bibfnamefont {S.}~\bibnamefont {Mancini}},\ }\bibfield  {title}
  {\bibinfo {title} {Quantum channels and memory effects},\ }\href
  {https://doi.org/10.1103/RevModPhys.86.1203} {\bibfield  {journal} {\bibinfo
  {journal} {Rev. Mod. Phys.}\ }\textbf {\bibinfo {volume} {86}},\ \bibinfo
  {pages} {1203} (\bibinfo {year} {2014})}\BibitemShut {NoStop}%
\bibitem [{\citenamefont {Wolf}(2011)}]{wolf2011url}%
  \BibitemOpen
  \bibfield  {author} {\bibinfo {author} {\bibfnamefont {M.~M.}\ \bibnamefont
  {Wolf}},\ }\href@noop {} {\emph {\bibinfo {title} {Quantum Channels and
  Operations-Guided Tour}}}\ (\bibinfo {year} {2011})\BibitemShut {NoStop}%
\bibitem [{\citenamefont {Watrous}(2018)}]{watrous2018theory}%
  \BibitemOpen
  \bibfield  {author} {\bibinfo {author} {\bibfnamefont {J.}~\bibnamefont
  {Watrous}},\ }\href@noop {} {\emph {\bibinfo {title} {The Theory of Quantum
  Information}}}\ (\bibinfo  {publisher} {{Cambridge university press}},\
  \bibinfo {year} {2018})\BibitemShut {NoStop}%
\bibitem [{\citenamefont {Oszmaniec}\ \emph {et~al.}(2017)\citenamefont
  {Oszmaniec}, \citenamefont {Guerini}, \citenamefont {Wittek},\ and\
  \citenamefont {Ac{\'i}n}}]{oszmaniec2017}%
  \BibitemOpen
  \bibfield  {author} {\bibinfo {author} {\bibfnamefont {M.}~\bibnamefont
  {Oszmaniec}}, \bibinfo {author} {\bibfnamefont {L.}~\bibnamefont {Guerini}},
  \bibinfo {author} {\bibfnamefont {P.}~\bibnamefont {Wittek}},\ and\ \bibinfo
  {author} {\bibfnamefont {A.}~\bibnamefont {Ac{\'i}n}},\ }\bibfield  {title}
  {\bibinfo {title} {Simulating {{Positive-Operator-Valued Measures}} with
  {{Projective Measurements}}},\ }\href
  {https://doi.org/10.1103/PhysRevLett.119.190501} {\bibfield  {journal}
  {\bibinfo  {journal} {Phys. Rev. Lett.}\ }\textbf {\bibinfo {volume} {119}},\
  \bibinfo {pages} {190501} (\bibinfo {year} {2017})}\BibitemShut {NoStop}%
\bibitem [{\citenamefont {Oszmaniec}\ \emph {et~al.}(2019)\citenamefont
  {Oszmaniec}, \citenamefont {Maciejewski},\ and\ \citenamefont
  {Pucha{\l}a}}]{oszmaniec2019}%
  \BibitemOpen
  \bibfield  {author} {\bibinfo {author} {\bibfnamefont {M.}~\bibnamefont
  {Oszmaniec}}, \bibinfo {author} {\bibfnamefont {F.~B.}\ \bibnamefont
  {Maciejewski}},\ and\ \bibinfo {author} {\bibfnamefont {Z.}~\bibnamefont
  {Pucha{\l}a}},\ }\bibfield  {title} {\bibinfo {title} {Simulating all quantum
  measurements using only projective measurements and postselection},\ }\href
  {https://doi.org/10.1103/PhysRevA.100.012351} {\bibfield  {journal} {\bibinfo
   {journal} {Phys. Rev. A}\ }\textbf {\bibinfo {volume} {100}},\ \bibinfo
  {pages} {012351} (\bibinfo {year} {2019})}\BibitemShut {NoStop}%
\bibitem [{\citenamefont {Singal}\ \emph {et~al.}(2022)\citenamefont {Singal},
  \citenamefont {Maciejewski},\ and\ \citenamefont {Oszmaniec}}]{singal2022}%
  \BibitemOpen
  \bibfield  {author} {\bibinfo {author} {\bibfnamefont {T.}~\bibnamefont
  {Singal}}, \bibinfo {author} {\bibfnamefont {F.~B.}\ \bibnamefont
  {Maciejewski}},\ and\ \bibinfo {author} {\bibfnamefont {M.}~\bibnamefont
  {Oszmaniec}},\ }\bibfield  {title} {\bibinfo {title} {Implementation of
  quantum measurements using classical resources and only a single ancillary
  qubit},\ }\href {https://doi.org/10.1038/s41534-022-00589-1} {\bibfield
  {journal} {\bibinfo  {journal} {npj Quantum Inf}\ }\textbf {\bibinfo {volume}
  {8}},\ \bibinfo {pages} {82} (\bibinfo {year} {2022})}\BibitemShut {NoStop}%
\bibitem [{\citenamefont {Linden}\ and\ \citenamefont
  {Skrzypczyk}(2023)}]{linden2023}%
  \BibitemOpen
  \bibfield  {author} {\bibinfo {author} {\bibfnamefont {N.}~\bibnamefont
  {Linden}}\ and\ \bibinfo {author} {\bibfnamefont {P.}~\bibnamefont
  {Skrzypczyk}},\ }\href@noop {} {\bibinfo {title} {How to use arbitrary
  measuring devices to perform almost perfect measurements}} (\bibinfo {year}
  {2023}),\ \Eprint {https://arxiv.org/abs/2203.02593} {arxiv:2203.02593
  [quant-ph]} \BibitemShut {NoStop}%
\bibitem [{\citenamefont {Bengtsson}\ and\ \citenamefont
  {{\.Z}yczkowski}(2017)}]{bengtsson2017geometry}%
  \BibitemOpen
  \bibfield  {author} {\bibinfo {author} {\bibfnamefont {I.}~\bibnamefont
  {Bengtsson}}\ and\ \bibinfo {author} {\bibfnamefont {K.}~\bibnamefont
  {{\.Z}yczkowski}},\ }\href@noop {} {\emph {\bibinfo {title} {Geometry of
  Quantum States: {{An}} Introduction to Quantum Entanglement}}}\ (\bibinfo
  {publisher} {{Cambridge university press}},\ \bibinfo {year}
  {2017})\BibitemShut {NoStop}%
\bibitem [{Note1()}]{Note1}%
  \BibitemOpen
  \bibinfo {note} {See Supplemental Material includes Ref. \cite
  {wolf2011url,watrous2018theory,PhysRevLett.116.240404,ma2016} at \protect
  \url {} for more details.}\BibitemShut {Stop}%
\bibitem [{\citenamefont {Arias}\ \emph {et~al.}(2002)\citenamefont {Arias},
  \citenamefont {Gheondea},\ and\ \citenamefont {Gudder}}]{arias2002fixed}%
  \BibitemOpen
  \bibfield  {author} {\bibinfo {author} {\bibfnamefont {A.}~\bibnamefont
  {Arias}}, \bibinfo {author} {\bibfnamefont {A.}~\bibnamefont {Gheondea}},\
  and\ \bibinfo {author} {\bibfnamefont {S.}~\bibnamefont {Gudder}},\
  }\bibfield  {title} {\bibinfo {title} {Fixed points of quantum operations},\
  }\href {https://doi.org/10.1063/1.1519669} {\bibfield  {journal} {\bibinfo
  {journal} {J. Math. Phys.}\ }\textbf {\bibinfo {volume} {43}},\ \bibinfo
  {pages} {5872} (\bibinfo {year} {2002})}\BibitemShut {NoStop}%
\bibitem [{\citenamefont {Burgarth}\ \emph {et~al.}(2013)\citenamefont
  {Burgarth}, \citenamefont {Chiribella}, \citenamefont {Giovannetti},
  \citenamefont {Perinotti},\ and\ \citenamefont {Yuasa}}]{Burgarth2013}%
  \BibitemOpen
  \bibfield  {author} {\bibinfo {author} {\bibfnamefont {D.}~\bibnamefont
  {Burgarth}}, \bibinfo {author} {\bibfnamefont {G.}~\bibnamefont
  {Chiribella}}, \bibinfo {author} {\bibfnamefont {V.}~\bibnamefont
  {Giovannetti}}, \bibinfo {author} {\bibfnamefont {P.}~\bibnamefont
  {Perinotti}},\ and\ \bibinfo {author} {\bibfnamefont {K.}~\bibnamefont
  {Yuasa}},\ }\bibfield  {title} {\bibinfo {title} {Ergodic and mixing quantum
  channels in finite dimensions},\ }\href
  {https://doi.org/10.1088/1367-2630/15/7/073045} {\bibfield  {journal}
  {\bibinfo  {journal} {New J. Phys.}\ }\textbf {\bibinfo {volume} {15}},\
  \bibinfo {pages} {073045} (\bibinfo {year} {2013})}\BibitemShut {NoStop}%
\bibitem [{\citenamefont {Albert}(2019)}]{Albert2019asymptoticsof}%
  \BibitemOpen
  \bibfield  {author} {\bibinfo {author} {\bibfnamefont {V.~V.}\ \bibnamefont
  {Albert}},\ }\bibfield  {title} {\bibinfo {title} {Asymptotics of quantum
  channels: {{Conserved}} quantities, an adiabatic limit, and matrix product
  states},\ }\href {https://doi.org/10.22331/q-2019-06-06-151} {\bibfield
  {journal} {\bibinfo  {journal} {Quantum}\ }\textbf {\bibinfo {volume} {3}},\
  \bibinfo {pages} {151} (\bibinfo {year} {2019})}\BibitemShut {NoStop}%
\bibitem [{Note2()}]{Note2}%
  \BibitemOpen
  \bibinfo {note} {Here we assume that the channel has no rotating points, see
  \cite {Note1} for the form of sequential channels that have rotating
  points.}\BibitemShut {Stop}%
\bibitem [{Note3()}]{Note3}%
  \BibitemOpen
  \bibinfo {note} {Considering that $\Tr (R_j)=0$ for decaying points and $\Tr
  (\rho _{\protect \rm fix}^i)=1$, so $\DOTSB \sum@ \slimits@ _{i=1}^n
  c_i+\DOTSB \sum@ \slimits@ _{j=n+1}^l c'_j=1$, and $c_1=1-\DOTSB \sum@
  \slimits@ _{i=2}^n c_i-\DOTSB \sum@ \slimits@ _{j=n+1}^l c'_j$}\BibitemShut
  {NoStop}%
\bibitem [{Note4()}]{Note4}%
  \BibitemOpen
  \bibinfo {note} {In order to use a single parameter $\gamma $ to adjust the
  relative strength, we assume that $B$ and $C$ have the same operator
  norm.}\BibitemShut {Stop}%
\bibitem [{\citenamefont {Stinespring}(1955)}]{stinespring}%
  \BibitemOpen
  \bibfield  {author} {\bibinfo {author} {\bibfnamefont {W.~F.}\ \bibnamefont
  {Stinespring}},\ }\bibfield  {title} {\bibinfo {title} {Positive functions on
  {{C}}*-{{Algebras}}},\ }\href@noop {} {\bibfield  {journal} {\bibinfo
  {journal} {Proc. Amer. Math. Soc.}\ }\textbf {\bibinfo {volume} {6}},\
  \bibinfo {pages} {211} (\bibinfo {year} {1955})}\BibitemShut {NoStop}%
\bibitem [{\citenamefont {Novotn\'{y}}\ \emph {et~al.}(2010)\citenamefont
  {Novotn\'{y}}, \citenamefont {Alber},\ and\ \citenamefont
  {Jex}}]{novotny2010asymptotic}%
  \BibitemOpen
  \bibfield  {author} {\bibinfo {author} {\bibfnamefont {J.}~\bibnamefont
  {Novotn\'{y}}}, \bibinfo {author} {\bibfnamefont {G.}~\bibnamefont {Alber}},\
  and\ \bibinfo {author} {\bibfnamefont {I.}~\bibnamefont {Jex}},\ }\bibfield
  {title} {\bibinfo {title} {Asymptotic evolution of random unitary
  operations},\ }\href {https://doi.org/doi:10.2478/s11534-010-0018-8}
  {\bibfield  {journal} {\bibinfo  {journal} {Open Physics}\ }\textbf {\bibinfo
  {volume} {8}},\ \bibinfo {pages} {1001} (\bibinfo {year} {2010})}\BibitemShut
  {NoStop}%
\bibitem [{\citenamefont {Audenaert}\ and\ \citenamefont
  {Scheel}(2008)}]{audenaert2008random}%
  \BibitemOpen
  \bibfield  {author} {\bibinfo {author} {\bibfnamefont {K.~M.~R.}\
  \bibnamefont {Audenaert}}\ and\ \bibinfo {author} {\bibfnamefont
  {S.}~\bibnamefont {Scheel}},\ }\bibfield  {title} {\bibinfo {title} {On
  random unitary channels},\ }\href
  {https://doi.org/10.1088/1367-2630/10/2/023011} {\bibfield  {journal}
  {\bibinfo  {journal} {New Journal of Physics}\ }\textbf {\bibinfo {volume}
  {10}},\ \bibinfo {pages} {023011} (\bibinfo {year} {2008})}\BibitemShut
  {NoStop}%
\bibitem [{\citenamefont {Bauer}\ and\ \citenamefont
  {Bernard}(2011)}]{bauer2011}%
  \BibitemOpen
  \bibfield  {author} {\bibinfo {author} {\bibfnamefont {M.}~\bibnamefont
  {Bauer}}\ and\ \bibinfo {author} {\bibfnamefont {D.}~\bibnamefont
  {Bernard}},\ }\bibfield  {title} {\bibinfo {title} {Convergence of repeated
  quantum nondemolition measurements and wave-function collapse},\ }\href
  {https://doi.org/10.1103/PhysRevA.84.044103} {\bibfield  {journal} {\bibinfo
  {journal} {Phys. Rev. A}\ }\textbf {\bibinfo {volume} {84}},\ \bibinfo
  {pages} {044103} (\bibinfo {year} {2011})}\BibitemShut {NoStop}%
\bibitem [{\citenamefont {Haapasalo}\ \emph {et~al.}(2016)\citenamefont
  {Haapasalo}, \citenamefont {Heinosaari},\ and\ \citenamefont
  {Kuramochi}}]{haapasalo2016}%
  \BibitemOpen
  \bibfield  {author} {\bibinfo {author} {\bibfnamefont {E.}~\bibnamefont
  {Haapasalo}}, \bibinfo {author} {\bibfnamefont {T.}~\bibnamefont
  {Heinosaari}},\ and\ \bibinfo {author} {\bibfnamefont {Y.}~\bibnamefont
  {Kuramochi}},\ }\bibfield  {title} {\bibinfo {title} {Saturation of repeated
  quantum measurements},\ }\href
  {https://doi.org/10.1088/1751-8113/49/33/33LT01} {\bibfield  {journal}
  {\bibinfo  {journal} {J. Phys. A: Math. Theor.}\ }\textbf {\bibinfo {volume}
  {49}},\ \bibinfo {pages} {33LT01} (\bibinfo {year} {2016})}\BibitemShut
  {NoStop}%
\bibitem [{\citenamefont {Ma}\ \emph {et~al.}(2018)\citenamefont {Ma},
  \citenamefont {Wang}, \citenamefont {Leong},\ and\ \citenamefont
  {Liu}}]{ma2018phase}%
  \BibitemOpen
  \bibfield  {author} {\bibinfo {author} {\bibfnamefont {W.-L.}\ \bibnamefont
  {Ma}}, \bibinfo {author} {\bibfnamefont {P.}~\bibnamefont {Wang}}, \bibinfo
  {author} {\bibfnamefont {W.-H.}\ \bibnamefont {Leong}},\ and\ \bibinfo
  {author} {\bibfnamefont {R.-B.}\ \bibnamefont {Liu}},\ }\bibfield  {title}
  {\bibinfo {title} {Phase transitions in sequential weak measurements},\
  }\href {https://doi.org/10.1103/PhysRevA.98.012117} {\bibfield  {journal}
  {\bibinfo  {journal} {Phys. Rev. A}\ }\textbf {\bibinfo {volume} {98}},\
  \bibinfo {pages} {012117} (\bibinfo {year} {2018})}\BibitemShut {NoStop}%
\bibitem [{\citenamefont {Ma}\ \emph {et~al.}(2023)\citenamefont {Ma},
  \citenamefont {Li},\ and\ \citenamefont {Liu}}]{PhysRevA.107.012217}%
  \BibitemOpen
  \bibfield  {author} {\bibinfo {author} {\bibfnamefont {W.-L.}\ \bibnamefont
  {Ma}}, \bibinfo {author} {\bibfnamefont {S.-S.}\ \bibnamefont {Li}},\ and\
  \bibinfo {author} {\bibfnamefont {R.-B.}\ \bibnamefont {Liu}},\ }\bibfield
  {title} {\bibinfo {title} {Sequential generalized measurements:
  {{Asymptotics}}, typicality, and emergent projective measurements},\ }\href
  {https://doi.org/10.1103/PhysRevA.107.012217} {\bibfield  {journal} {\bibinfo
   {journal} {Phys. Rev. A}\ }\textbf {\bibinfo {volume} {107}},\ \bibinfo
  {pages} {012217} (\bibinfo {year} {2023})}\BibitemShut {NoStop}%
\bibitem [{\citenamefont {Yang}\ \emph {et~al.}(2017)\citenamefont {Yang},
  \citenamefont {Ma},\ and\ \citenamefont {Liu}}]{yang2017}%
  \BibitemOpen
  \bibfield  {author} {\bibinfo {author} {\bibfnamefont {W.}~\bibnamefont
  {Yang}}, \bibinfo {author} {\bibfnamefont {W.-L.}\ \bibnamefont {Ma}},\ and\
  \bibinfo {author} {\bibfnamefont {R.-B.}\ \bibnamefont {Liu}},\ }\bibfield
  {title} {\bibinfo {title} {Quantum many-body theory for electron spin
  decoherence in nanoscale nuclear spin baths},\ }\href
  {https://doi.org/10.1088/0034-4885/80/1/016001} {\bibfield  {journal}
  {\bibinfo  {journal} {Rep. Prog. Phys.}\ }\textbf {\bibinfo {volume} {80}},\
  \bibinfo {pages} {016001} (\bibinfo {year} {2017})}\BibitemShut {NoStop}%
\bibitem [{\citenamefont {Sza{\'n}kowski}\ \emph {et~al.}(2017)\citenamefont
  {Sza{\'n}kowski}, \citenamefont {Ramon}, \citenamefont {Krzywda},
  \citenamefont {Kwiatkowski} \emph {et~al.}}]{szankowski2017}%
  \BibitemOpen
  \bibfield  {author} {\bibinfo {author} {\bibfnamefont {P.}~\bibnamefont
  {Sza{\'n}kowski}}, \bibinfo {author} {\bibfnamefont {G.}~\bibnamefont
  {Ramon}}, \bibinfo {author} {\bibfnamefont {J.}~\bibnamefont {Krzywda}},
  \bibinfo {author} {\bibfnamefont {D.}~\bibnamefont {Kwiatkowski}}, \emph
  {et~al.},\ }\bibfield  {title} {\bibinfo {title} {Environmental noise
  spectroscopy with qubits subjected to dynamical decoupling},\ }\href@noop {}
  {\bibfield  {journal} {\bibinfo  {journal} {Journal of Physics: Condensed
  Matter}\ }\textbf {\bibinfo {volume} {29}},\ \bibinfo {pages} {333001}
  (\bibinfo {year} {2017})}\BibitemShut {NoStop}%
\bibitem [{\citenamefont {Liu}\ \emph {et~al.}(2012)\citenamefont {Liu},
  \citenamefont {Pan}, \citenamefont {Jiang}, \citenamefont {Zhao},\ and\
  \citenamefont {Liu}}]{liu2012}%
  \BibitemOpen
  \bibfield  {author} {\bibinfo {author} {\bibfnamefont {G.-Q.}\ \bibnamefont
  {Liu}}, \bibinfo {author} {\bibfnamefont {X.-Y.}\ \bibnamefont {Pan}},
  \bibinfo {author} {\bibfnamefont {Z.-F.}\ \bibnamefont {Jiang}}, \bibinfo
  {author} {\bibfnamefont {N.}~\bibnamefont {Zhao}},\ and\ \bibinfo {author}
  {\bibfnamefont {R.-B.}\ \bibnamefont {Liu}},\ }\bibfield  {title} {\bibinfo
  {title} {Controllable effects of quantum fluctuations on spin free-induction
  decay at room temperature},\ }\href@noop {} {\bibfield  {journal} {\bibinfo
  {journal} {Scientific reports}\ }\textbf {\bibinfo {volume} {2}},\ \bibinfo
  {pages} {432} (\bibinfo {year} {2012})}\BibitemShut {NoStop}%
\bibitem [{\citenamefont {Reinhard}\ \emph {et~al.}(2012)\citenamefont
  {Reinhard}, \citenamefont {Shi}, \citenamefont {Zhao}, \citenamefont {Rempp},
  \citenamefont {Naydenov}, \citenamefont {Meijer}, \citenamefont {Hall},
  \citenamefont {Hollenberg}, \citenamefont {Du}, \citenamefont {Liu},\ and\
  \citenamefont {Wrachtrup}}]{Reinhard2012}%
  \BibitemOpen
  \bibfield  {author} {\bibinfo {author} {\bibfnamefont {F.}~\bibnamefont
  {Reinhard}}, \bibinfo {author} {\bibfnamefont {F.}~\bibnamefont {Shi}},
  \bibinfo {author} {\bibfnamefont {N.}~\bibnamefont {Zhao}}, \bibinfo {author}
  {\bibfnamefont {F.}~\bibnamefont {Rempp}}, \bibinfo {author} {\bibfnamefont
  {B.}~\bibnamefont {Naydenov}}, \bibinfo {author} {\bibfnamefont
  {J.}~\bibnamefont {Meijer}}, \bibinfo {author} {\bibfnamefont {L.~T.}\
  \bibnamefont {Hall}}, \bibinfo {author} {\bibfnamefont {L.}~\bibnamefont
  {Hollenberg}}, \bibinfo {author} {\bibfnamefont {J.}~\bibnamefont {Du}},
  \bibinfo {author} {\bibfnamefont {R.-B.}\ \bibnamefont {Liu}},\ and\ \bibinfo
  {author} {\bibfnamefont {J.}~\bibnamefont {Wrachtrup}},\ }\bibfield  {title}
  {\bibinfo {title} {Tuning a spin bath through the quantum-classical
  transition},\ }\href {https://doi.org/10.1103/PhysRevLett.108.200402}
  {\bibfield  {journal} {\bibinfo  {journal} {Phys. Rev. Lett.}\ }\textbf
  {\bibinfo {volume} {108}},\ \bibinfo {pages} {200402} (\bibinfo {year}
  {2012})}\BibitemShut {NoStop}%
\bibitem [{\citenamefont {Plenio}\ and\ \citenamefont
  {Knight}(1998)}]{plenio1998quantum}%
  \BibitemOpen
  \bibfield  {author} {\bibinfo {author} {\bibfnamefont {M.~B.}\ \bibnamefont
  {Plenio}}\ and\ \bibinfo {author} {\bibfnamefont {P.~L.}\ \bibnamefont
  {Knight}},\ }\bibfield  {title} {\bibinfo {title} {The quantum-jump approach
  to dissipative dynamics in quantum optics},\ }\href
  {https://doi.org/10.1103/RevModPhys.70.101} {\bibfield  {journal} {\bibinfo
  {journal} {Rev. Mod. Phys.}\ }\textbf {\bibinfo {volume} {70}},\ \bibinfo
  {pages} {101} (\bibinfo {year} {1998})}\BibitemShut {NoStop}%
\bibitem [{\citenamefont {Brun}(2002)}]{brun2002simple}%
  \BibitemOpen
  \bibfield  {author} {\bibinfo {author} {\bibfnamefont {T.~A.}\ \bibnamefont
  {Brun}},\ }\bibfield  {title} {\bibinfo {title} {{A simple model of quantum
  trajectories}},\ }\href {https://doi.org/10.1119/1.1475328} {\bibfield
  {journal} {\bibinfo  {journal} {Am. J. Phys.}\ }\textbf {\bibinfo {volume}
  {70}},\ \bibinfo {pages} {719} (\bibinfo {year} {2002})}\BibitemShut
  {NoStop}%
\bibitem [{\citenamefont {Wudarski}\ \emph {et~al.}(2023)\citenamefont
  {Wudarski}, \citenamefont {Zhang},\ and\ \citenamefont
  {Dykman}}]{Wudarski2023}%
  \BibitemOpen
  \bibfield  {author} {\bibinfo {author} {\bibfnamefont {F.}~\bibnamefont
  {Wudarski}}, \bibinfo {author} {\bibfnamefont {Y.}~\bibnamefont {Zhang}},\
  and\ \bibinfo {author} {\bibfnamefont {M.~I.}\ \bibnamefont {Dykman}},\
  }\bibfield  {title} {\bibinfo {title} {Nonergodic measurements of qubit
  frequency noise},\ }\href {https://doi.org/10.1103/PhysRevLett.131.230201}
  {\bibfield  {journal} {\bibinfo  {journal} {Phys. Rev. Lett.}\ }\textbf
  {\bibinfo {volume} {131}},\ \bibinfo {pages} {230201} (\bibinfo {year}
  {2023})}\BibitemShut {NoStop}%
\bibitem [{\citenamefont {Doherty}\ \emph {et~al.}(2012)\citenamefont
  {Doherty}, \citenamefont {Dolde}, \citenamefont {Fedder}, \citenamefont
  {Jelezko}, \citenamefont {Wrachtrup}, \citenamefont {Manson},\ and\
  \citenamefont {Hollenberg}}]{PhysRevB.85.205203}%
  \BibitemOpen
  \bibfield  {author} {\bibinfo {author} {\bibfnamefont {M.~W.}\ \bibnamefont
  {Doherty}}, \bibinfo {author} {\bibfnamefont {F.}~\bibnamefont {Dolde}},
  \bibinfo {author} {\bibfnamefont {H.}~\bibnamefont {Fedder}}, \bibinfo
  {author} {\bibfnamefont {F.}~\bibnamefont {Jelezko}}, \bibinfo {author}
  {\bibfnamefont {J.}~\bibnamefont {Wrachtrup}}, \bibinfo {author}
  {\bibfnamefont {N.~B.}\ \bibnamefont {Manson}},\ and\ \bibinfo {author}
  {\bibfnamefont {L.~C.~L.}\ \bibnamefont {Hollenberg}},\ }\bibfield  {title}
  {\bibinfo {title} {Theory of the ground-state spin of the {{NV}}{$^-$} center
  in diamond},\ }\href {https://doi.org/10.1103/PhysRevB.85.205203} {\bibfield
  {journal} {\bibinfo  {journal} {Phys. Rev. B}\ }\textbf {\bibinfo {volume}
  {85}},\ \bibinfo {pages} {205203} (\bibinfo {year} {2012})}\BibitemShut
  {NoStop}%
\bibitem [{\citenamefont {Zhao}\ \emph {et~al.}(2012)\citenamefont {Zhao},
  \citenamefont {Ho},\ and\ \citenamefont {Liu}}]{PhysRevB.85.115303}%
  \BibitemOpen
  \bibfield  {author} {\bibinfo {author} {\bibfnamefont {N.}~\bibnamefont
  {Zhao}}, \bibinfo {author} {\bibfnamefont {S.-W.}\ \bibnamefont {Ho}},\ and\
  \bibinfo {author} {\bibfnamefont {R.-B.}\ \bibnamefont {Liu}},\ }\bibfield
  {title} {\bibinfo {title} {Decoherence and dynamical decoupling control of
  nitrogen vacancy center electron spins in nuclear spin baths},\ }\href
  {https://doi.org/10.1103/PhysRevB.85.115303} {\bibfield  {journal} {\bibinfo
  {journal} {Phys. Rev. B}\ }\textbf {\bibinfo {volume} {85}},\ \bibinfo
  {pages} {115303} (\bibinfo {year} {2012})}\BibitemShut {NoStop}%
\bibitem [{\citenamefont {Viola}\ \emph {et~al.}(1999)\citenamefont {Viola},
  \citenamefont {Knill},\ and\ \citenamefont {Lloyd}}]{viola1999}%
  \BibitemOpen
  \bibfield  {author} {\bibinfo {author} {\bibfnamefont {L.}~\bibnamefont
  {Viola}}, \bibinfo {author} {\bibfnamefont {E.}~\bibnamefont {Knill}},\ and\
  \bibinfo {author} {\bibfnamefont {S.}~\bibnamefont {Lloyd}},\ }\bibfield
  {title} {\bibinfo {title} {Dynamical {{Decoupling}} of {{Open Quantum
  Systems}}},\ }\href {https://doi.org/10.1103/PhysRevLett.82.2417} {\bibfield
  {journal} {\bibinfo  {journal} {Phys. Rev. Lett.}\ }\textbf {\bibinfo
  {volume} {82}},\ \bibinfo {pages} {2417} (\bibinfo {year}
  {1999})}\BibitemShut {NoStop}%
\bibitem [{\citenamefont {Ryan}\ \emph {et~al.}(2010)\citenamefont {Ryan},
  \citenamefont {Hodges},\ and\ \citenamefont {Cory}}]{ryan2010}%
  \BibitemOpen
  \bibfield  {author} {\bibinfo {author} {\bibfnamefont {C.~A.}\ \bibnamefont
  {Ryan}}, \bibinfo {author} {\bibfnamefont {J.~S.}\ \bibnamefont {Hodges}},\
  and\ \bibinfo {author} {\bibfnamefont {D.~G.}\ \bibnamefont {Cory}},\
  }\bibfield  {title} {\bibinfo {title} {Robust {{Decoupling Techniques}} to
  {{Extend Quantum Coherence}} in {{Diamond}}},\ }\href
  {https://doi.org/10.1103/PhysRevLett.105.200402} {\bibfield  {journal}
  {\bibinfo  {journal} {Phys. Rev. Lett.}\ }\textbf {\bibinfo {volume} {105}},\
  \bibinfo {pages} {200402} (\bibinfo {year} {2010})}\BibitemShut {NoStop}%
\bibitem [{\citenamefont {Ma}\ and\ \citenamefont {Liu}(2016)}]{ma2016}%
  \BibitemOpen
  \bibfield  {author} {\bibinfo {author} {\bibfnamefont {W.-L.}\ \bibnamefont
  {Ma}}\ and\ \bibinfo {author} {\bibfnamefont {R.-B.}\ \bibnamefont {Liu}},\
  }\bibfield  {title} {\bibinfo {title} {Angstrom-{{Resolution Magnetic
  Resonance Imaging}} of {{Single Molecules}} via {{Wave-Function
  Fingerprints}} of {{Nuclear Spins}}},\ }\href
  {https://doi.org/10.1103/PhysRevApplied.6.024019} {\bibfield  {journal}
  {\bibinfo  {journal} {Phys. Rev. Applied}\ }\textbf {\bibinfo {volume} {6}},\
  \bibinfo {pages} {024019} (\bibinfo {year} {2016})}\BibitemShut {NoStop}%
\bibitem [{\citenamefont {Blume-Kohout}\ \emph {et~al.}(2008)\citenamefont
  {Blume-Kohout}, \citenamefont {Ng}, \citenamefont {Poulin},\ and\
  \citenamefont {Viola}}]{Blume-Kohout2008}%
  \BibitemOpen
  \bibfield  {author} {\bibinfo {author} {\bibfnamefont {R.}~\bibnamefont
  {Blume-Kohout}}, \bibinfo {author} {\bibfnamefont {H.~K.}\ \bibnamefont
  {Ng}}, \bibinfo {author} {\bibfnamefont {D.}~\bibnamefont {Poulin}},\ and\
  \bibinfo {author} {\bibfnamefont {L.}~\bibnamefont {Viola}},\ }\bibfield
  {title} {\bibinfo {title} {Characterizing the structure of preserved
  information in quantum processes},\ }\href
  {https://doi.org/10.1103/PhysRevLett.100.030501} {\bibfield  {journal}
  {\bibinfo  {journal} {Phys. Rev. Lett.}\ }\textbf {\bibinfo {volume} {100}},\
  \bibinfo {pages} {030501} (\bibinfo {year} {2008})}\BibitemShut {NoStop}%
\bibitem [{\citenamefont {Blume-Kohout}\ \emph {et~al.}(2010)\citenamefont
  {Blume-Kohout}, \citenamefont {Ng}, \citenamefont {Poulin},\ and\
  \citenamefont {Viola}}]{Blume-Kohout2010}%
  \BibitemOpen
  \bibfield  {author} {\bibinfo {author} {\bibfnamefont {R.}~\bibnamefont
  {Blume-Kohout}}, \bibinfo {author} {\bibfnamefont {H.~K.}\ \bibnamefont
  {Ng}}, \bibinfo {author} {\bibfnamefont {D.}~\bibnamefont {Poulin}},\ and\
  \bibinfo {author} {\bibfnamefont {L.}~\bibnamefont {Viola}},\ }\bibfield
  {title} {\bibinfo {title} {Information-preserving structures: A general
  framework for quantum zero-error information},\ }\href
  {https://doi.org/10.1103/PhysRevA.82.062306} {\bibfield  {journal} {\bibinfo
  {journal} {Phys. Rev. A}\ }\textbf {\bibinfo {volume} {82}},\ \bibinfo
  {pages} {062306} (\bibinfo {year} {2010})}\BibitemShut {NoStop}%
\end{thebibliography}%


\begin{thebibliography}{5}%
\makeatletter
\providecommand \@ifxundefined [1]{%
 \@ifx{#1\undefined}
}%
\providecommand \@ifnum [1]{%
 \ifnum #1\expandafter \@firstoftwo
 \else \expandafter \@secondoftwo
 \fi
}%
\providecommand \@ifx [1]{%
 \ifx #1\expandafter \@firstoftwo
 \else \expandafter \@secondoftwo
 \fi
}%
\providecommand \natexlab [1]{#1}%
\providecommand \enquote  [1]{``#1''}%
\providecommand \bibnamefont  [1]{#1}%
\providecommand \bibfnamefont [1]{#1}%
\providecommand \citenamefont [1]{#1}%
\providecommand \href@noop [0]{\@secondoftwo}%
\providecommand \href [0]{\begingroup \@sanitize@url \@href}%
\providecommand \@href[1]{\@@startlink{#1}\@@href}%
\providecommand \@@href[1]{\endgroup#1\@@endlink}%
\providecommand \@sanitize@url [0]{\catcode `\\12\catcode `\$12\catcode
  `\&12\catcode `\#12\catcode `\^12\catcode `\_12\catcode `\%12\relax}%
\providecommand \@@startlink[1]{}%
\providecommand \@@endlink[0]{}%
\providecommand \url  [0]{\begingroup\@sanitize@url \@url }%
\providecommand \@url [1]{\endgroup\@href {#1}{\urlprefix }}%
\providecommand \urlprefix  [0]{URL }%
\providecommand \Eprint [0]{\href }%
\providecommand \doibase [0]{https://doi.org/}%
\providecommand \selectlanguage [0]{\@gobble}%
\providecommand \bibinfo  [0]{\@secondoftwo}%
\providecommand \bibfield  [0]{\@secondoftwo}%
\providecommand \translation [1]{[#1]}%
\providecommand \BibitemOpen [0]{}%
\providecommand \bibitemStop [0]{}%
\providecommand \bibitemNoStop [0]{.\EOS\space}%
\providecommand \EOS [0]{\spacefactor3000\relax}%
\providecommand \BibitemShut  [1]{\csname bibitem#1\endcsname}%
\let\auto@bib@innerbib\@empty
\bibitem [{\citenamefont {Wolf}(2011)}]{wolf2011url}%
  \BibitemOpen
  \bibfield  {author} {\bibinfo {author} {\bibfnamefont {M.~M.}\ \bibnamefont
  {Wolf}},\ }\href@noop {} {\emph {\bibinfo {title} {Quantum Channels and
  Operations-Guided Tour}}}\ (\bibinfo {year} {2011})\BibitemShut {NoStop}%
\bibitem [{Note1()}]{Note1}%
  \BibitemOpen
  \bibinfo {note} {See proposition 6.2 in \cite {wolf2011url}}\BibitemShut
  {NoStop}%
\bibitem [{\citenamefont {Watrous}(2018)}]{watrous2018theory}%
  \BibitemOpen
  \bibfield  {author} {\bibinfo {author} {\bibfnamefont {J.}~\bibnamefont
  {Watrous}},\ }\href@noop {} {\emph {\bibinfo {title} {The Theory of Quantum
  Information}}}\ (\bibinfo  {publisher} {{Cambridge university press}},\
  \bibinfo {year} {2018})\BibitemShut {NoStop}%
\bibitem [{\citenamefont {Macieszczak}\ \emph {et~al.}(2016)\citenamefont
  {Macieszczak}, \citenamefont {Gu{\c t}{\u a}}, \citenamefont {Lesanovsky},\
  and\ \citenamefont {Garrahan}}]{PhysRevLett.116.240404}%
  \BibitemOpen
  \bibfield  {author} {\bibinfo {author} {\bibfnamefont {K.}~\bibnamefont
  {Macieszczak}}, \bibinfo {author} {\bibfnamefont {M.}~\bibnamefont {Gu{\c
  t}{\u a}}}, \bibinfo {author} {\bibfnamefont {I.}~\bibnamefont
  {Lesanovsky}},\ and\ \bibinfo {author} {\bibfnamefont {J.~P.}\ \bibnamefont
  {Garrahan}},\ }\bibfield  {title} {\bibinfo {title} {Towards a theory of
  metastability in open quantum dynamics},\ }\href
  {https://doi.org/10.1103/PhysRevLett.116.240404} {\bibfield  {journal}
  {\bibinfo  {journal} {Phys. Rev. Lett.}\ }\textbf {\bibinfo {volume} {116}},\
  \bibinfo {pages} {240404} (\bibinfo {year} {2016})}\BibitemShut {NoStop}%
\bibitem [{\citenamefont {Ma}\ and\ \citenamefont {Liu}(2016)}]{ma2016}%
  \BibitemOpen
  \bibfield  {author} {\bibinfo {author} {\bibfnamefont {W.-L.}\ \bibnamefont
  {Ma}}\ and\ \bibinfo {author} {\bibfnamefont {R.-B.}\ \bibnamefont {Liu}},\
  }\bibfield  {title} {\bibinfo {title} {Angstrom-{{Resolution Magnetic
  Resonance Imaging}} of {{Single Molecules}} via {{Wave-Function
  Fingerprints}} of {{Nuclear Spins}}},\ }\href
  {https://doi.org/10.1103/PhysRevApplied.6.024019} {\bibfield  {journal}
  {\bibinfo  {journal} {Phys. Rev. Applied}\ }\textbf {\bibinfo {volume} {6}},\
  \bibinfo {pages} {024019} (\bibinfo {year} {2016})}\BibitemShut {NoStop}%
\end{thebibliography}%
\newpage

\end{document}


\title{Supplementary Material for "Theory of Metastability in Discrete-Time Open Quantum Dynamics"}
\author{Yuan-De Jin}\thanks{These authors contributed equally to this work}
\affiliation{Department of Applied Physics,University of Science and Technology Beijing, Beijing 100083, China}
\author{Chu-Dan Qiu}\thanks{These authors contributed equally to this work}
\affiliation{Laboratory of Semiconductor Physics, Institute of Semiconductors, Chinese Academy of Sciences, Beijing, 100083, China}
\author{Wen-Long Ma}
\email{wenlongma@semi.ac.cn}
\affiliation{Laboratory of Semiconductor Physics, Institute of Semiconductors, Chinese Academy of Sciences, Beijing, 100083, China}
\affiliation{Center of Materials Science and Opto-Electronic Technology,\\ University of Chinese Academy of Sciences, Beijing 100049, China}

\date{\today}
\begin{abstract}
	In this supplementary material, we first provide more details about fundamentals of quantum channels, and properties of fixed points and extreme metastable states (EMSs) in Ramsey interferometry measurements (RIMs). Then we present Monte Carlo simulation results for a target system composed of multiple qubits, with the ancilla qubit under both RIMs and dynamical decoupling (DD) sequences. Moreover, we numerically demonstrate that quantum metastability is robust even when the target system suffers additional dissipation.
\end{abstract}
\maketitle

\tableofcontents

\section{Fundamentals of quantum channels}
\subsection{Definition of quantum channels}
Denote the linear operators acting on a Hilbert space $\h$ as $\bh$, then a map $\Phi:\bh\to\mathcal{B(H)}$ is a quantum channel if it satisfies the following conditions \cite{wolf2011url}
\begin{itemize}
  \item \textbf{Linear map}: For any $A,B\in\bh$ and complex number $c$, $\Phi(A+cB)=\Phi(A)+cT(B)$;
  \item \textbf{Trace preserving}: For any $A\in \bh$, $\Tr[\Phi(A)]=\Tr(A)$. This implies unitality of $\Phi^{\dagger}$, i.e., $\Phi^{\dagger}(\I)=\I$, where $\Phi^{\dagger}$ is defined by $\Tr[B\Phi(A)]=\Tr[\Phi^{\dagger}(B)A]$;
  \item \textbf{Completely positive}: For any positive operator $\rho$, $(\Phi\otimes \I)(\rho)$ is still a positive operator, with $\I$ is the identity operator on an additional system with dimension ${\rm dim} (\h)$.
\end{itemize}
So a quantum channel is a completely positive and trace-preserving (CPTP) linear map, which maps a state to another one by $\Phi(\rho)=\rho'$.

\subsection{Representations of quantum channels}
Every quantum channel has four different representations: the Kraus representation, the Stinespring representation, the natural representation, and the Choi representation. In this paper, we use the first three representations of quantum channels.

\subsubsection{Kraus representation}
Of the four representations, the Kraus representation is the most commonly used one. In this representation, a quantum channel is fully characterized by a collection of Kraus operators $\{M_\alpha\}_{\alpha=1}^r$ satisfying $\sum_\alpha M^{\dagger}_{\alpha}M_{\alpha} =\mathbb{I}$ so that
\begin{equation}\label{phisup}
  \Phi(\cdot)=\sum_{\alpha=1}^r M_{\alpha} (\cdot) M_{\alpha}^\dagger=\sum_{\alpha=1}^r \mathcal{M}_{\alpha}(\cdot),
\end{equation}
where $\mathcal{M_{\alpha}}(\cdot)=M_{\alpha} (\cdot) M_{\alpha}^{\dagger}$ is a superoperator.

\subsubsection{Stinespring representation}
The stinespring representation is a dilation of a quantum channel. The dilation can be realized by coupling the target system to an ancilla system, and letting the composite system undergoing a unitary evolution and then tracing over the ancilla system,
\begin{equation}
  \Phi(\rho)=\Tr_a[U(\rho_a\otimes\rho)U^\dagger],
\end{equation}
where $\rho_a$ is the initial state of the ancilla system and $U$ is a unitary of the composite system and ${\rm Tr}_a$ denotes the trace over the ancilla system. For a $r$-dimensional ancilla system with an orthonormal basis $\{|\alpha\}_{\alpha=1}^r$ and the initial state $\rho_a=|1\rangle_a\langle 1|$, the Kraus operator can be easily obtained as $M_{\alpha}=\langle r|U|1\rangle_a$.

\subsubsection{Natural representation}
For sequential applications of the same channel, it is convenient to use the natural representation of a quantum channel. To understand this representation, we need to introduce Hilbert-Schmidt (HS) space. In the HS space, an operator on a $d$-dimensional Hilbert space (represented by a $d\times d$ matrix) is transformed to a $d^2\times 1$ vector,
 \begin{equation}
  A=\mqty[a_{11}&\dots& a_{1d}\\\vdots&\ddots&\vdots\\a_{d1}&\dots& a_{dd}]=\mqty[\vb*{a}_1\\\vdots \\ \vb*{a}_d] \Rightarrow \kett{A}=\mqty[\,\vb*{a}^T_1\,\\\vdots \\\, \vb*{a}^T_d\,],
\end{equation}
where the superscript $T$ denotes the matrix transposition. Such a transformation can also be represented by $A=\sum_{ij}a_{ij}\ket{i} \bra{j}$$\Rightarrow$$\kett{A}=\sum_{ij}a_{ij}\kett{ij}$ with $\kett{ij}=\ket{i}\otimes \ket{j}$. The inner product in the HS space is defined as
\begin{equation}
\begin{aligned}
	    \brakett{A}{B}&=\sum_{ijpq}a^*_{ij}b_{pq}\brakett{ij}{pq}\\
	    &=\sum_{ij}a^*_{ij}b_{ij}\\
	    &=\sum_{j}(A^\dagger B)_{jj}\\
	    &=\Tr(A^\dagger B).
\end{aligned}\end{equation}
Then a superoperator (assuming $\mathcal{O}(\rho)=A\rho B$) becomes a single matrix $A\otimes B^T$ acting on a vector $\kett{\rho}$ in HS space,
\begin{equation}
  A\rho B=\sum_{ijpq}a_{ij}\rho_{jk}b_{pq}\ket{i}\bra{q} \,\,\Rightarrow\,\, A\otimes B^T\kett{\rho},
\end{equation}
so the natural representation of the channel in Eq. (\ref{phisup}) in HS space is
 \begin{equation}
   \hat{\Phi}=\sum_{\alpha=1}^r \hat{\mathcal{M}}_\alpha
\end{equation}
where $\hat{\mathcal{M}}_\alpha=M_\alpha\otimes M^*_\alpha$. Note that we add hats on operators acting on HS space.

\subsection{Jordan decomposition of a quantum channel}
The natural representation of a quantum channel on the HS space is a $d^2\times d^2$ square operator. This operator may not be diagonalizable, but can always be converted to a Jordan normal form as
\begin{equation}
  \begin{aligned}
	  \phii&=S\left(\bigoplus_{k=1}^\kappa \mathcal{J}_{d_k}(\lambda_k)\right)S^{-1}	\\
	  &=S\left(\sum_{\left|{\lambda}_j\right|=1} {\lambda}_j {\mathcal{P}}_j+ \sum_{\left|{\lambda}_k\right|<1} ({\mathcal{P}}_k+{\mathcal{N}}_k)\right)S^{-1},
\end{aligned}
\end{equation}
where $S$ is an invertible $d^2\times d^2$ matrix, $\kappa$ is the total number of Jordan blocks, and $\mathcal{J}_{d_i}(\lambda_i)$ represents a $d_i$-dimensional Jordan block corresponding to the eigenvalue $\lambda_i$, ${\mathcal{P}}_j$ is a projection operator and ${\mathcal{N}}_k$ is a nilpotent operator satisfying ${\mathcal{N}}_k^{d_k}=0$. Considering that $\sum_k^\kappa d_k=d$, then the channel is diagonalizable iff $\kappa=d$. Note that the Jordan blocks corresponding to the fixed points or rotating points (with eigenvalues $\abs{\lambda_i}=1$) are all rank-one projectors \footnote{See proposition 6.2 in \cite{wolf2011url}}.

If the channel is diagonalizable, we have
\begin{equation}
\begin{aligned}
	    \hat\Phi&=\sum_i \lambda_i |R_i\rangle\rangle \langle\langle L_i|,	
\end{aligned}\label{phidecomp}
\end{equation}
where $\{|R_i\rangle\rangle, |L_i\rangle\rangle\}$ is a complete biorthogonal basis satisfying $\brakett{L_i}{R_j}=\delta_{ij}$. The trace-preserving property of $\hat{\Phi}$ implies the unitality of $\hat{\Phi}^\dagger$, i.e., $\hat \Phi^\dagger\kett{\mathbb{I}}=\kett{\I}$ or there exists a left eigenvector $\langle\langle L_i|=\langle\langle\mathbb{I}|$ for eigenvalue 1, then for right eigenvectors corresponding to $\abs{\lambda_j}<1$, we have
\begin{equation}
  0=\brakett{L_i}{R_j}={\Tr}(L_i^\dagger R_j)=\Tr(R_j),
\end{equation}

\subsection{Sequential quantum channels with rotating points}
If the channel has $n$ eigenvectors corresponding to eigenvalue $\abs{\lambda}=1$, in which there are $n_0$ fixed points and $n-n_0$ rotating points with $\lambda_j=\ee^{\ii \varphi_j}$, and $l-n$ eigenvectors with $\lambda_k=\abs{\lambda_k}\ee^{\ii \varphi_k}\approx 1$, then after sequentially applying the quantum channel for $m$ times, we have
\begin{equation}
\begin{aligned}
	    \hat\Phi^m\kett{\rho}&=\sum_{i=1}^{n_0} c_i\kett{\rho_{\rm fix}^i}+\sum_{j=n_0+1}^n c_j\lambda_j^m|{R_{\rm rot}^j}\rangle\rangle+\sum_{k=n+1}^{l}c_k\lambda_k^m \kett{R_k}+\cdots	\\
	    &\simeq\sum_{i=1}^n c_i\kett{\rho_{\rm fix}^i}+\sum_{j=n_0+1}^n c_j\ee^{\ii m\varphi_j}|{R_{\rm rot}^j}\rangle\rangle\sum_{k=n+1}^{l}c_k\ee^{ m(\ln{\abs{\lambda_k}+\ii \varphi_k)}} \kett{R_k},
\end{aligned}
\end{equation}
where we truncate the equation as we did in the main text.

Then, in the metastable region, $\ee^{ m\ln{\abs{\lambda_k}}}\approx 1$, by considering conjugate pairs of complex eigenvalues and absorbing the phase in $c$, we have
\begin{equation}\label{channelmeta}
   \hat\Phi^m\kett{\rho}\simeq\sum_{i=1}^{n_0} c_i \kett{\rho_{\rm fix}^i}+\sum_{j=n_0+1}^{l}c_j'(m)\kett{R_j'},
\end{equation}
which is approximately the same as Eq.(4) in the main text, except that the summation of metastable points also include the rotating points.

\section{Fixed points of the channel induced by RIMs}

Now we consider the fixed points of quantum channel on a target system induced by a RIM of an ancilla qubit. The anciila is coupled to the target system with a pure-dephasing Hamiltonian,
\begin{equation}\label{hamiltonian}
  H=\sigma_q^z\otimes B+ \gamma \,\mathbb{I}_q\otimes C,
\end{equation}
where $\sigma_q^i$ is the Pauli-$i$ operator of the ancilla, $B$ and $C$ are both operators on the target system and $\gamma$ controls the magnitude of the second term. For a RIM, the Kraus representation of the channel on the target system is
\begin{equation}
  \phii=\mm_0+\mm_1=(\uu_0+\uu_1)/2,\label{RIMKraus}
\end{equation}
where $\mathcal{\hat U_\alpha}=U_\alpha\otimes U_\alpha^*$ with $U_{\alpha}=e^{-i[(-1)^{\alpha} B+\gamma C]}$,
and $\mathcal{\hat M_\alpha}=M_\alpha\otimes M_\alpha^*$ with the Kraus operator $M_{\alpha}=[U_0-(-1)^{\alpha} e^{i\Delta\phi}U_1]/2$ and $\Delta\phi=\phi_1-\phi_2$.
Then the fixed points induced by such a unital channel is given by the following proposition.

\textbf{Proposition 1}. The fixed points of the channel in Eq. (\ref{RIMKraus}) depends on the commutativity of $B$ and $C$. If $[B,C]=0$, the fixed points are spanned by a set of rank-one projections $\{|j\rangle\langle j|\}_{j=1}^d$; if $[B,C]\neq 0$, the fixed points are spanned by a set of projection operators $\{\Pi_j\}_{j=1}^r$ ($r\leq d$), satisfying $\sum_{j=1}^r\Pi_j=\mathbb{I}$.

\begin{proof}
It has been proven that $\rho$ is a fixed point of a unital channel if and only if it commutes with every Kraus operator \cite{watrous2018theory}, i.e., $[\rho, M_{\alpha}]=0$ for any $\alpha$. This implies that $[\rho, U_0]=[\rho,U_1]=0$. If the above condition is always satisfied for any $\alpha$, then $[\rho,B]=[\rho,C]=0$.

If $[B,C]=0$, then $B$ and $C$ can be diagonalized simultaneously, $B=\sum_{j=1}^d b_j \ket{j}\bra{j}$ and $C=\sum_{j=1}^d c_j \ket{j}\bra{j}$. So the fixed points must include the rank-one projections $\{|j\rangle\langle j|\}_{j=1}^d$ and their linear combinations.

If $[B,C]\neq 0$, we can block diagonalize them simultaneously by unitary transformation,
\begin{equation}
  B=W\left(\bigoplus_{j=1}^r B_j\right) W^{\dagger}, \quad C=W\left(\bigoplus_{j=1}^r C_j\right) W^{\dagger}
\end{equation}
where $r\leq d$ is the number of blocks (with equality occurring only when $[B,C]=0$ and all of blocks are one-dimensional), $W$ is unitary matrix and should be chosen so that $B_j$ and $C_j$ for any $j$ cannot be reduced further to have more blocks. There must be at least one subspace $\h_j$ in which $[B_j,C_j]\neq 0$ to make $[B,C]\neq 0$. Such a block diagonalization partitions the Hilbert space of the target system into the direct sum of $r$ subspaces $\mathcal{H}=\bigoplus_{j=1}^r\mathcal{H}_j$, and $[B_j, {C}_{j}] \neq 0$ for at least one subspace $\mathcal{H}_j$ with ${\rm dim}(\mathcal{H}_j)\geq2$. Thus the Kraus operator is also transformed to a block-diagonal form as $M_\alpha=\oplus_{j=1}^r M_\alpha^j$. Then the fixed points must include the set of projections $\{\Pi_j\}_{j=1}^r$ ($r\leq d$) and their linear combinations, where $\Pi_j=WP_jW^{-1}$ with $P_j$ being the projector to $\mathcal{H}_j$. Note that the case $[B,C]=0$ can be regarded as a special case of $[B,C]\neq 0$.

Now we prove that there are no other fixed points for the case $[B,C]\neq 0$, where there is at least one block with $[B_j,C_j]\neq 0$ and $[M_0^j,M_1^j]\neq0$.
Suppose there is another density matrix satisfying $[\rho',M_\alpha^j]=0$. If $\rank{\rho'}=d_j$, then $[\rho',M_0^j]=[\rho',M_1^j]=0$. Since the positive operator $\rho'$ can be diagonalized, this implies that $[M_0^j,M_1^j]=0$. If $\rank(\rho')<d_j$, then formulate another fixed point $\rho''=\rho'+\eta\mathbb{I}$ with $\eta$ being a positive number such that $\rank(\rho'')=d_j$, then the proof is similar to the former case.

\end{proof}

\section{Quantum metastability in sequential RIMs for a target qubit}
\subsection{Construction of EMSs}
If target system is a single qubit with $B=\sigma_z$ and $C=0$, the fixed points are spanned by $\kett{00}$ and $\kett{11}$, where $\{|0\rangle, |1\rangle\}$ are eigenstates of $\sigma_z$. When there exists a small perturbing Hamiltonian $\gamma C=\gamma \sigma_x$, the channel has a single fixed point $\kett{\rho_{\rm fix}}=\kett{\mathbb I}/2$, and another metastable point $\kett{R_2}$.
So the dimension of MM is $2-1=1$. In the metastable region, $\lambda_2^m\approx 1$ (assuming $\lambda_2$ is real), then a metastable state is
\begin{equation}
  \kett{\rho_{\rm MS}}=\kett{\rho_{\rm fix}}+c_2\kett{R_2},
\end{equation}
with $c_2={\Tr}(L_2^\dagger \rho)$. Thus the metastable state is determined only by $c_2$, and the extremal points in the one-dimensional MM are
\begin{equation}
  \kett{\rho_1}=\kett{\rho_{\rm fix}}+c_2^M/h\kett{R_2},\quad  \kett{\rho_2}=\kett{\rho_{\rm fix}}+c_2^m/h\kett{R_2},
\end{equation}
where $c_2^{M}$ ($c_2^{m}$) is the maximal (minimal) value of $c_2$, and $h=\sqrt{\brakett{L_2}{L_2}\brakett{R_2}{R_2}}$ is a normalization coefficient that will be derived below. For real $\lambda_2$, $R_2$ and $L_2$ can be chosen to be Hermitian \cite{PhysRevLett.116.240404}, then $c_2=\Tr(\rho L_2)$ and
$c_2^{M}$ and $c_2^{m}$ are two eigenvalues of $L_2$. So we have $L_2=c_2^M \ket{M}\bra{M}+c_2^m \ket{m}\bra{m}$ with $\{\ket{M},\ket{m}\}$ forming a complete orthonormal basis.

Let $\rho=k_1 \ket{M}\bra{M}+k_2 \ket{m}\bra{m}$ with $k_1+k_2=1$, then $\rho L_2=c_2^M k_1 \ket{M}\bra{M}+c_2^m k_2 \ket{m}\bra{m}$. This means that only when $k_1=1,\, k_2=0$ (or $\rho=\ket{M}\bra{M}$), we have $c_2=c_2^M$. Since $\rho_{1,2}$ should be pure states, this means $\Tr(\rho_{1,2}^2)=1$,
\begin{equation}
\begin{aligned}
	    1=\Tr(\rho_1^2)&=\Tr(\mathbb{I}^2/4+c_2^M/h R_2+(c_2^M/h)^2 R_2^2)\\
	    &=1/2+(c_2^M/h)^2 \Tr(R_2^2)\\
        &\Downarrow \\
	    1/2&=(c_2^M/h)^2 \Tr(R_2^2).
\end{aligned}\end{equation}
We can derive a similar expression for $\rho_2$, i.e., $1/2=(c_2^m/h)^2 \Tr(R_2^2)$, so
\begin{equation}
\begin{aligned}
	     h^2=[(c_2^M)^2+(c_2^m)^2 ]\Tr(R_2^2)	
	     =\Tr(L_2^2)\Tr(R_2^2)
	     =\brakett{L_2}{L_2}\brakett{R_2}{R_2}.
\end{aligned}
\end{equation}

We can illustrate the above analysis with a simple example. Consider a fixed point $\kett{\rho_{\rm fix}}=\kett{\mathbb{I}}/2=(\kett{00}+\kett{11})/2$, and a metastable point with $\kett{R_2}=(\kett{00}-\kett{11})/2$, then $\kett{L_2}=\kett{00}-\kett{11}$, where $c_2^M=1,\,c_2^m=-1$, and $h=\sqrt{\brakett{L_2}{L_2}\brakett{R_2}{R_2}}=1$. Thus the EMSs are
\begin{equation}
  |\rho_1\rangle\rangle=|\rho_{\rm fix}\rangle\rangle+|R_2\rangle\rangle=|00\rangle\rangle \langle\langle00|,\quad  |\rho_2\rangle\rangle=|\rho_{\rm fix}\rangle\rangle-|R_2\rangle\rangle=|11\rangle\rangle \langle\langle11|,
\end{equation}
For general cases, the EMSs are the eigenstates of $B$ up to some corrections.

\subsection{Construction of metastable states}
Any metastable state can be represented by a mixture of EMSs,
\begin{equation}
  \kett{\rho_{\rm MS}}=p_1\kett{\rho_1}+p_2\kett{\rho_2}
\end{equation}
where $p_{1,2}=\brakett{P_1}{\rho}=\Tr(\rho P_{1,2})$ with $P_{1,2}$ being observables satisfying $\langle\langle{P_v}|{\rho_u}\rangle\rangle=\delta_{v,u}$, $P_{1}+P_2=\mathbb{I}$ and $P_{1,2}\geq 0$. Below we show how to construct such observables.

Since $\kett{\rho_{\rm MS}}=\Phi^m \kett{\rho}$ in the metastable region $\mu''\ll m \ll \mu'$, so we have
\begin{equation}
  \Phi^m\simeq|\rho_1\rangle\rangle\langle\langle P_1|+|\rho_2\rangle\rangle\langle\langle P_2|,
\end{equation}
Comparing the above equation with $\Phi^m\simeq \frac{1}{2}|\mathbb{I}\rangle\rangle\langle\langle\mathbb{I}|+|R_2\rangle\rangle\langle\langle L_2|$ in metastable region, we have
\begin{equation}
  \begin{aligned}
	\frac{1}{2}|\mathbb{I}\rangle\rangle \langle\langle\mathbb{I}|+|R_2\rangle\rangle \langle\langle L_2| &=  \left(\frac{1}{2}|\mathbb{I}\rangle\rangle+\frac{c_2^M}{h}|R_2\rangle\rangle\right)\langle\langle P_1|+\left(\frac{1}{2}|\mathbb{I}\rangle\rangle+\frac{c_2^m}{h}|R_2\rangle\rangle\right)\langle\langle P_2|	\\
    &\Downarrow \\
	|R_2\rangle\rangle \langle\langle L_2|&=\frac{c_2^M}{h}|R_2\rangle\rangle \langle\langle P_1|+\frac{c_2^m}{h}|R_2\rangle\rangle(\langle\langle\mathbb{I}|-\langle\langle P_1|)\\
	 &=|R_2\rangle\rangle\left[\frac{c_2^M}{h}\langle\langle P_1|+\frac{c_2^m}{h}(\langle\langle\mathbb{I}|-\langle\langle P_1|)\right],
\end{aligned}
\end{equation}
so we have
\begin{equation}
  P_1=\frac{hL_2-c_2^m \mathbb{I}}{\Delta c_2},\quad P_2= \frac{-hL_2+c_2^M \mathbb{I}}{\Delta c_2},
\end{equation}
with $\Delta c_2=c_2^M-c_2^m$.


\section{Quantum metastability in sequential quantum channels for multiple target qubits}
In this section, we consider a practical example to show quantum metastability in sequential quantum channels, that is, an NV center electron spin (ancilla qubit) in a $^{13}$C nuclear spins bath (target system). The coupling Hamiltonian has a form similar to Eq. (\ref{hamiltonian}),
\begin{equation}
  H=\sigma_q^z\otimes B+C,
\end{equation}
with
\begin{equation}
  B=f(t)\sum_{k=1}^K \vb*{A}_k\cdot \vb*{I}_k, \quad C=\omega_L\sum_{k=1}^K I_k^z+\sum_{k<j} D_{jk}\left[\boldsymbol{I}_k \cdot \boldsymbol{I}_j-\frac{3\left(\boldsymbol{I}_k \cdot \boldsymbol{r}_{k j}\right)\left(\boldsymbol{I}_j \cdot \boldsymbol{r}_{k j}\right)}{r_{k j}^2}\right],
\end{equation}
where $f(t)$ is a modulation function accounting for possible DD control of the ancilla qubit, $\vb*{A}_k=(A_k^{x}, A_k^{y}, A_k^{z})$ is the hyperfine interaction vector of the $k$${\rm th}$ nuclear spin, $\vb*{I}_k$ is the nuclear spin operator vector, $d=2^K$ is the dimension of Hilbert space of target system, $\omega_L=\gamma_n B_z$ is Larmor precession frequency of nuclear spins, $D_{jk}=\frac{\mu_0 \gamma_{\mathrm{n}}^2}{4 \pi r_{k j}^3}$ with $\gamma_n$ is the gyromagnetic ratios of the target spins, $\vb*{r}_{kj}$ is the displacement from the $i$${\rm th}$ target spin to the $j$${\rm th}$ target spin and $\mu_0$ is vacuum permeability. Note that we let $\gamma=1$ and use practical parameters below.

\subsection{RIM sequences}

For RIM sequences, we assume that the external magnetic field is zero or very weak, so that
\begin{equation}
  B=\sum_{k=1}^K \vb*{A}_k\cdot \vb*{I}_k, \quad C=\sum_{k<j} D_{jk}\left[\boldsymbol{I}_k \cdot \boldsymbol{I}_j-\frac{3\left(\boldsymbol{I}_k \cdot \boldsymbol{r}_{k j}\right)\left(\boldsymbol{I}_j \cdot \boldsymbol{r}_{k j}\right)}{r_{k j}^2}\right],
\end{equation}

\begin{figure}[htbp]
\centering
  \includegraphics[width=16cm]{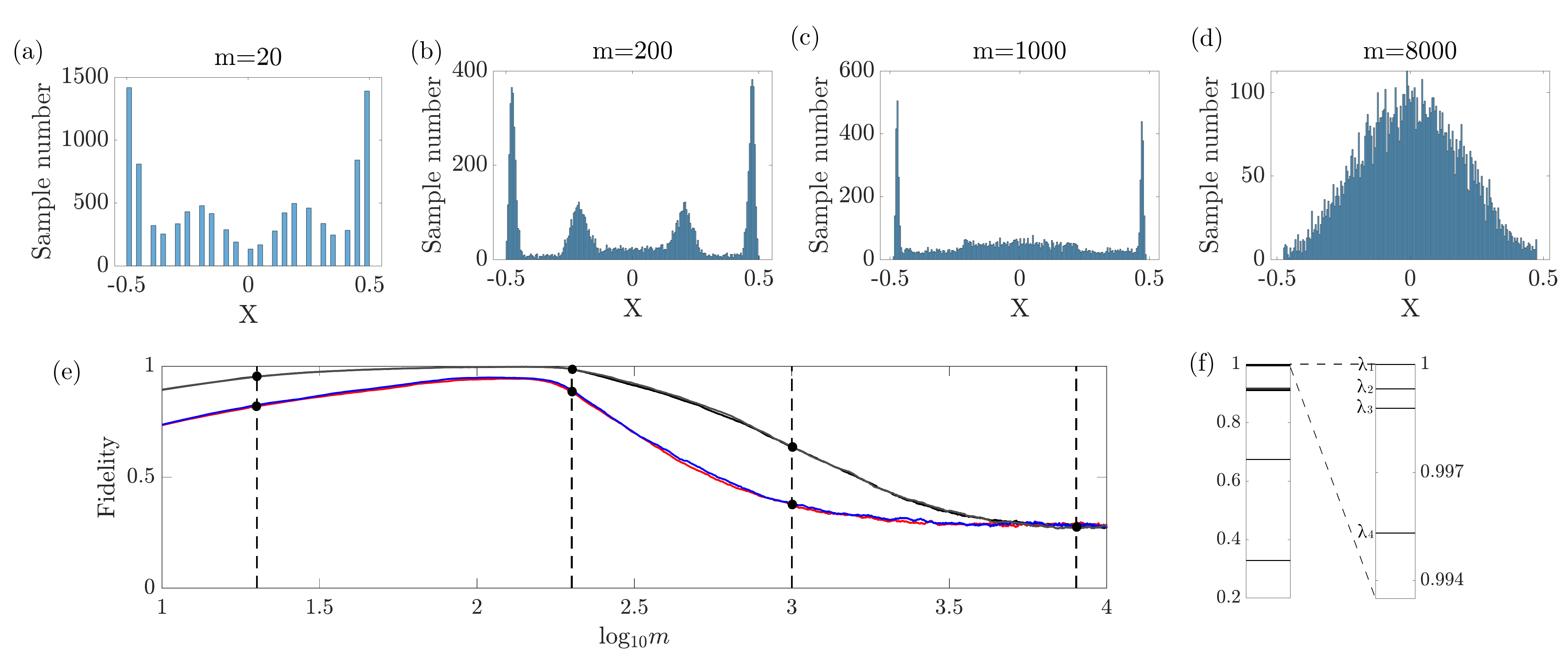}
  \caption{Metastability in measurement statistics of sequence RIMs. (a-d)The Monte Carlo simulation for two target qubits, in which we present four stages of the system evolution. (a) When the time of measurements is small, the width of peaks are relatively large and the peaks are not absolutely distinguishable. (b) For larger $m$, the four peaks corresponding to four EMSs are clear. (c) The EMSs relevant to $\kett{22}$ and $\kett{33}$ vanish faster, with that relevant to $\kett{11}$ and $\kett{44}$ are still existing. (d) All EMSs are collapsed. (e) The trajectory of evolution of fidelity, in which the stages in (a-d) are labeled by dashed line. In all Monte Carlo simulations, we use $10^4$ samples, parameters are $\Delta \phi=\frac{\pi}{2}$, $\abs{A_1}=0.585\mathrm{kHz},\, \abs{A_2}=0.890\mathrm{kHz}, \abs{D_{12}}=11.6\mathrm{Hz}$, then $||H_e||/||V||=0.0316$. (f) Spectra of the quantum channel. }\label{doublequbit}
\end{figure}

\begin{figure}[htbp]
\centering
  \includegraphics[width=16cm]{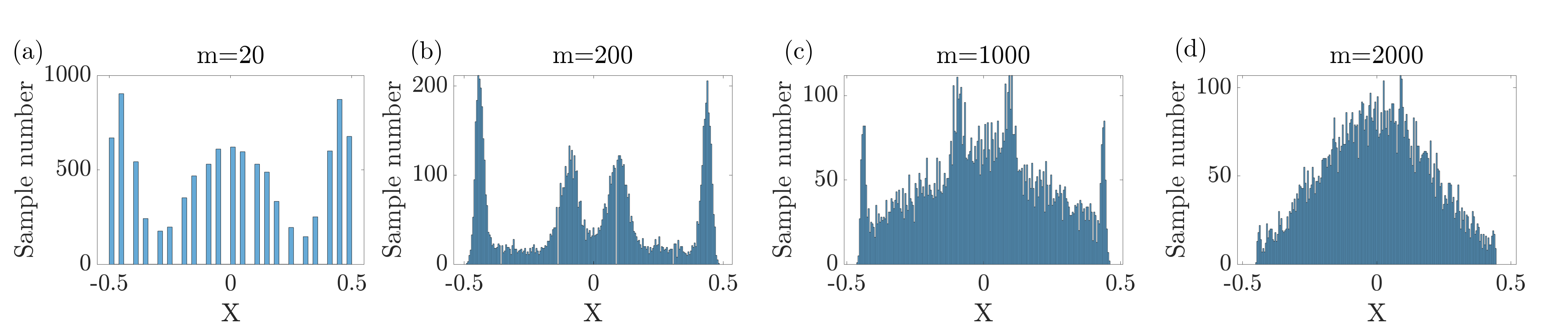}
  \caption{Monte Carlo simulation for two target qubits. The dynamics of polarization and depolarization can be seen from (a) to (d), which proves that our metastability theory also applies to the system involving Zeeman term. In all Monte Carlo simulation, we use $10^4$ samples, $\Delta \phi=\frac{\pi}{2}$ is chosen.  $\abs{A_1}=0.138\mathrm{MHz},\, \abs{A_2}=0.517\mathrm{MHz}, \abs{D_{12}}=60.5\mathrm{Hz}$, $B=20$G, then $\omega_L=13.5$ KHz, and $||H_e||/||V||=0.0412$. }\label{doublequbitmag}
\end{figure}

To show metastable polarization of nuclear spins, we also measure the evolution of fidelity, which is defined as
\begin{equation}
  F_{i}(\rho)=\Tr\sqrt{\sqrt{\rho_{i}}\rho\sqrt{\rho_{i}}},
\end{equation}
where $\rho_{i}=\ket{i}\bra{i}$ with $\ket{i}$ being the eigenstate of $B$ with $i\in\{1,2,3,4\}$. Fidelity here can show the distance between the state of system and unperturbed eigenstates. We can see the target system are polarized to nearby with the unperturbed eigenstates, which are our EMSs, and the fidelity plateaus can also confirm the metastability behaviors [Fig. \ref{doublequbit}]. When there is an external magnetic field involving, metastability also remains if the field is weak (Fig. \ref{doublequbitmag}), i.e., $\gamma_n B\ll \abs{A_k}$.


\subsection{DD sequences}

\begin{figure}[htbp]
\centering
  \includegraphics[width=10cm]{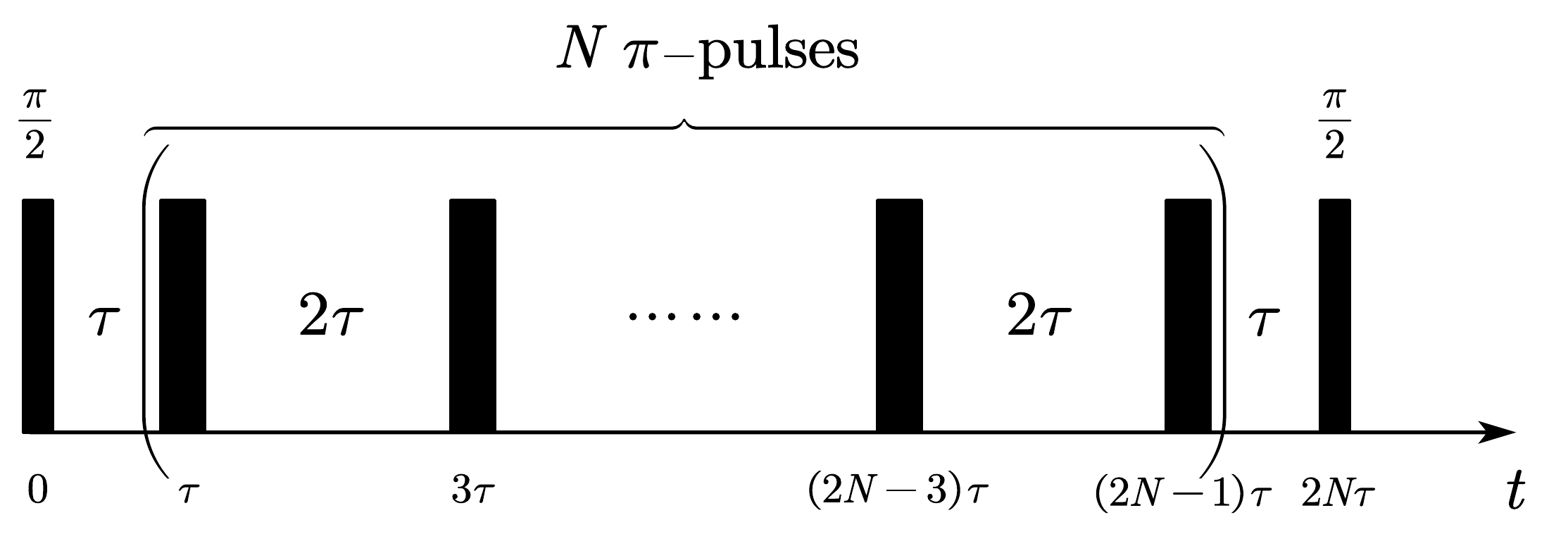}
  \caption{Schematic of an $N$-pulse CPMG sequence with $N$ $\pi-$flips at $t=\tau,3\tau,\cdots,(2N-1)\tau$.}\label{DDS}
\end{figure}

DD sequences are a generalization of RIM sequences, with additional $N$ $\pi$-flips of the ancilla spin during each cycle [Fig. \ref{DDS}]. Then we have
\begin{equation}
  B=f(t)\sum_{k=1}^K \vb*{A}_k\cdot \vb*{I}_k, \quad C=\omega_L\sum_{k=1}^K I_k^z,
\end{equation}
where $f(t)$ is the DD modulation function jumping between $+1$ and $-1$ every time the sensor is flipped by a DD pulse. Here for simplicity, we neglect the dipolar interaction term in $C$, which means these target spins are spatially far away from each other.

For $N$-pulse Carr-Purcell-Meiboom-Gill (CPMG) control, $f(t)=f(t+T)$ with $T=4\tau$. Specifically, $f(t)=-1$ when $\tau+\beta T\leq t<3\tau+\beta T$ with $\beta=0,1,\dots,\frac{N}{2}-1$ and $f(t)=1$ otherwise. Then $f(t)$ can be expanded into a Fourier series as $f(t)=\sum_{n=1}^\infty c_n \cos(n\omega_T t)$, where $c_n$ is $n$th Fourier expansion coefficient and $\omega_T=\frac{2\pi}{4\tau}$ is the angular frequency. For the weak coupling condition, i.e., $\abs{\vb*{A}_k}\ll \omega_L$, then we can approximate $f(t)$ by keeping only the first term in the Fourier series, $f(t)\approx c_1\cos(\omega_T t)=\frac{4}{\pi}\cos(\omega_T t)$.
Then $B$ becomes,
\begin{equation}
  B=\frac{4}{\pi} \cos(\omega_T t)\sum_{k=1}^K \vb*{A}_k\cdot \vb*{I}_k\approx\frac{4}{\pi} \cos(\omega_T t)\sum_{k=1}^K A^\perp_k I^\perp_k,
\end{equation}
where $A^\perp_k=\sqrt{{A_k^x}^2+{A_k^y}^2}$ and $I_k^\perp=\cos\alpha I_k^x+\sin\alpha I_k^y$, and the longitude component $A_z I_z$ can be neglected in the weak coupling region \cite{ma2016}.

\begin{figure}[H]
\centering
  \includegraphics[width=18cm]{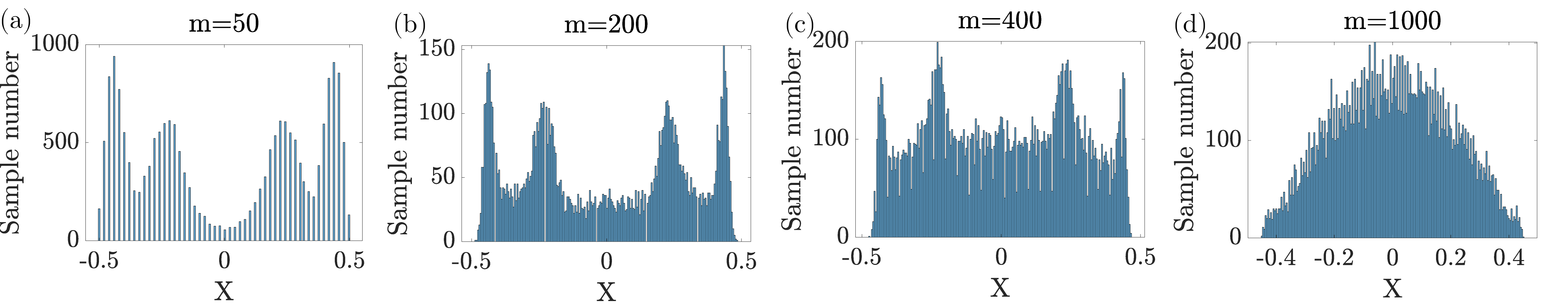}
  \caption{Monte Carlo simulation for two target spins with DD sequences and growing time of measurements $m$. As $m$ grows, the peaks emerge and become apparent (a-b) since they are corresponding to the four EMSs, then they vanish gradually as $m$ approaching the boundary of metastable region (c), which finally collapse to the maximally mixed state (d). In all Monte Carlo simulation, we use $10^4$ samples, parameters are $\Delta \phi=\frac{\pi}{2}$, $\abs{A_1}=5\mathrm{kHz},\, \abs{A_2}=6\mathrm{kHz}$, $B=200$G, $\Delta_\omega=10^{-3}\omega_L$, $N=32$.}
  \label{DDfig}
\end{figure}

Now we move to the rotating frame with respect to $\omega_T I_z$, by using rotating wave approximation, the effective Hamiltonian becomes time-independent,
\begin{equation}\label{DDHfinal}
   B=\frac{2}{\pi} \sum_{k=1}^K A^\perp_k I^\perp_k,\quad C=\Delta_{\omega}\sum_{k=1}^K I_k^z,
\end{equation}
where $\Delta_{\omega}=\omega_L-\omega_T$ is the detuning away from Larmor precession frequency.
Then we can see from Eq.(\ref{DDHfinal}) that, if we choose $\omega_T=\omega_L$, also called resonance condition, then $C=0$ and the target qubits can be polarized. However, practically the adjustment of frequency may not be precise. With a small detuning, quantum metastability theory shows that the target can still be polarized with appropriate rounds of DD, as numerically verified in Fig. \ref{DDfig}.

\section{Quantum metastability for dissipative target system}

In this section, we show that quantum metastability is robust even when the target system suffers additional dissipations. Suppose that the target system (a single qubit) suffers dephasing and relaxation noise, the evolution of the composite systems can be described by the following Lindblad master equation,
\begin{equation}
  \dv{\rho_{\rm tot}}{t}=-i[H,\rho_{\rm tot}]+\sum_k \Gamma_k \left (L_k\rho_{\rm tot} L_k^{\dagger}-\frac{1}{2}\left\{L_k^{\dagger} L_k,\rho_{\rm tot}\right\}\right),
\end{equation}
where $\rho_{\rm tot}$ is the density matrix of the composite system, $H=\sigma_z\otimes B+ \gamma \,\mathbb{I}_q\otimes C$ , $L_1=\sigma_z$ denotes the target dephasing, $L_2=\sigma^-=|1\rangle_q\langle0|$ denotes the target relaxation, and $\Gamma_k$ is the dissipation rate.

\begin{figure}[htbp]
\centering
  \includegraphics[width=18cm]{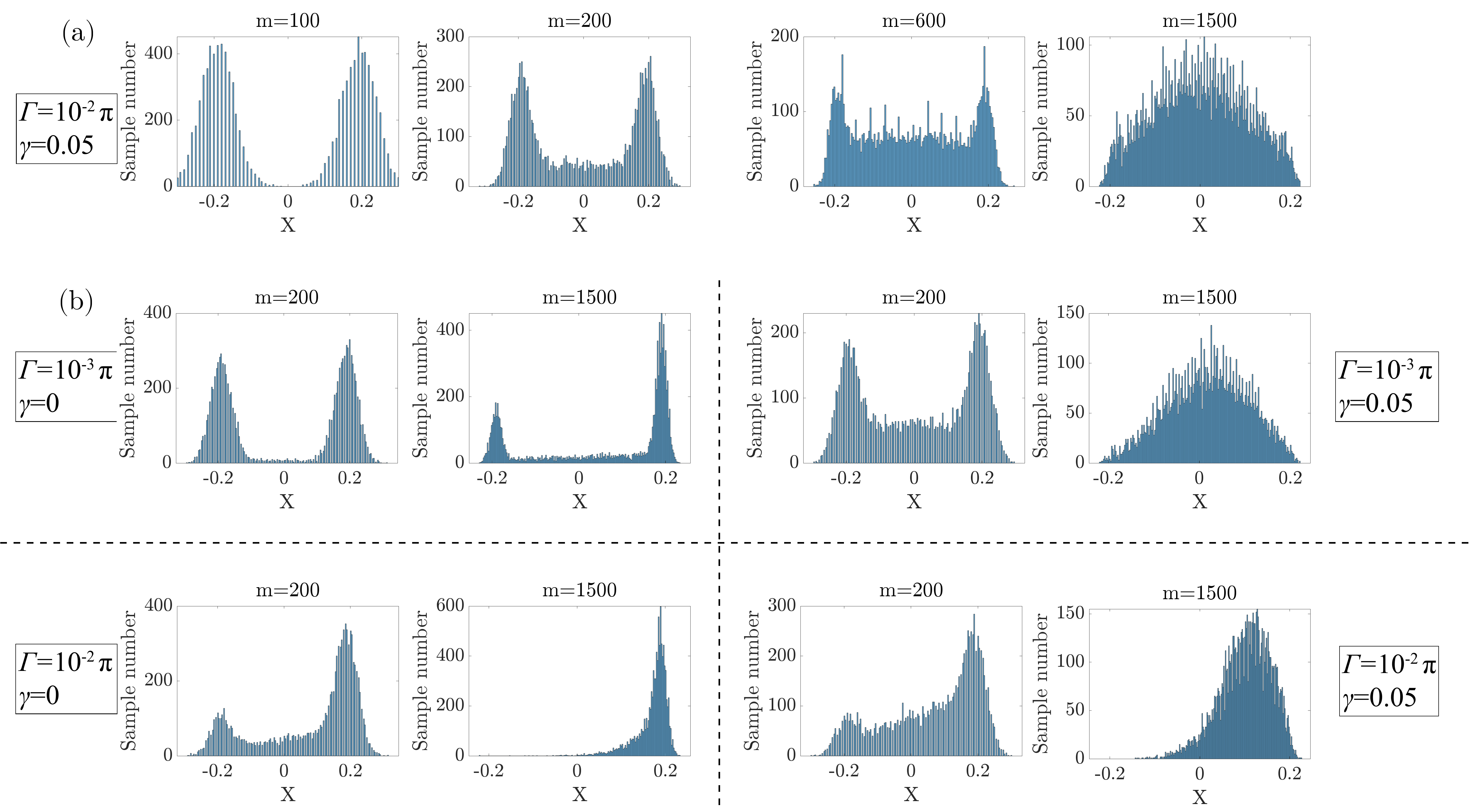}
  \caption{Monte Carlo simulations for a target qubit with (a) dephasing noise and (b) relaxation noise. (a) is like Fig. 2 in our main text, showing that there is no influence in our discussion.  In (b) it can be seen that the dissipation of the target system remains independent of the conclusion of the metastable, both for the case of $\gamma = 0$ (left panel) and for $\gamma = 0.05$ (right panel), the peak on the left side is gradually vanished as the round of the measurements increasing due to $\sigma^-$ makes $\ket{0}$ into $\ket{1}$. In all Monte Carlo simulation, we use $10^4$ samples, $\Delta\phi=\frac{\pi}{2}$ is chosen.}
  \label{diss}
\end{figure}

We perform Monte Carlo simulations for each kind of dissipation separately, in order to examine their influence. The results show that the dephasing noise of target system does not influence the measurement statistics. However, the relaxation of target system maps $\ket{0}$ to $\ket{1}$, which makes the peak corresponding to $\kett{00}$ (when $\gamma=0$) or that corresponding to EMS perturbed from $\kett{00}$ (when $\gamma\neq 0$) transfer to the other peak. This does not influence our discussion about metastability, if the dissipate rate is very small or when $m$ is relatively small (see upper panel of Fig. \ref{diss}{\color{blue}(b)}).

\bibliography{Metastability}
